\documentclass[12pt]{article}

\usepackage{subfig}
\usepackage{color,colortbl}
\usepackage{epsfig,amssymb,amsfonts,amsmath,graphicx,dsfont,cite,xfrac}
\usepackage{authblk}
\usepackage{rotating}
\usepackage{graphicx}
\usepackage{multirow}
\usepackage{tikz,pgfplots}
\usetikzlibrary{decorations.pathreplacing}
\usetikzlibrary{shapes.multipart}
\usetikzlibrary{positioning}
\usetikzlibrary{matrix}

\usepackage{cite}
\usepackage[colorlinks=true,linkcolor=blue,citecolor=red]{hyperref}
\title{Third-order ladder operators, generalized Okamoto and exceptional orthogonal polynomials}

\author[${1,2}$]{V. Hussin\thanks{veronique.hussin@umontreal.ca}}
\author[${3}$]{I. Marquette\thanks{i.marquette@uq.edu.au}}
\author[${2}$]{K. Zelaya\thanks{Corresponding author: zelayame@crm.umontreal.ca}}

\affil[${1}$]{\footnotesize Centre de Recherches Math\'ematiques, Universit\'e de Montr\'eal, Montr\'eal H3C 3J7, QC, Canada}

\affil[${2}$]{\footnotesize D\'epartement de Math\'ematiques et de Statistique, Universit\'e de Montr\'eal, Montr\'eal H3C 3J7, QC, Canada}

\affil[${3}$]{\footnotesize School of Mathematics and Physics, The University of Queensland, Brisbane, QLD 4072, Australia}

\date{}

\begin{document}

\maketitle

\begin{abstract}
We extend and generalize the construction of Sturm-Liouville problems for a family of Hamiltonians constrained to fulfill a third-order shape-invariance condition and focusing on the ``$-2x/3$'' hierarchy of solutions to the fourth Painlev\'e transcendent. Such a construction has been previously addressed in the literature for some particular cases but we realize it here in the most general case. The corresponding potential in the Hamiltonian operator is a rationally extended oscillator defined in terms of the conventional Okamoto polynomials, from which we identify three different zero-modes constructed in terms of the generalized Okamoto polynomials. The third-order ladder operators of the system reveal that the complete set of eigenfunctions is decomposed as a union of three disjoint sequences of solutions, generated from a set of three-term recurrence relations. We also identify a link between the eigenfunctions of the Hamiltonian operator and a special family of exceptional Hermite polynomial.

\end{abstract}

%%%%%%%%%%%%%%%%%%%%%%%%%%%%%%%%%%%%%%%%%%%%%%%

%---------------------------------------> Section
\section{Introduction}
Nonlinear equations have played a fundamental role in understanding the dynamics of some physical models, even in cases where the governing physical laws are defined in terms of linear equations~\cite{Sch18}. Particularly, in quantum mechanics, the nonlinear Riccati equation~\cite{Inc56,Sch18} has found several interesting applications in the study of exactly-solvable models. Such a task is achieved through the factorization method~\cite{Sch40a,Sch40b,Sch41,Inf51,Mie84,Mie00,Mie04}, a technique intrinsically related to the Darboux transformation~\cite{Dar88,Mat91}. The latter is also known as \textit{supersymmetric quantum mechanics} (SUSY QM) because of the mathematical equivalence with the supersymmetric construction of Witten~\cite{Wit85,Coo01,Jun96} in the potential theory model. For stationary quantum systems, that is, time-independent Hamiltonians, the corresponding dynamical law reduces to a Sturm-Liouville eigenvalue problem in the form of second-order differential equation in the spatial coordinates, the solutions of which can be addressed in terms of either hypergeometric or confluent hypergeometric functions~\cite{Mil98,Der11} for some particular interactions such as the harmonic oscillator, hydrogen atom, interaction between diatomic (Morse potential) and polyatomic molecules (Rosen-Morse potential), transparent potential interactions (P\"oschl-teller potential), among others. In this regard, the Darboux transformation has led to an outstanding progress in the study of exactly solvable quantum models, where the models previously mentioned can be used as a departing point to generate new families of potentials with spectrum on demand. Moreover, the latter does not restrict to Hermitian Hamiltonians, and non-Hermitian models can be constructed after imposing the non-self-adjoint condition on the resulting Hamiltonians~\cite{Can98,Ros15}.  The realitiy of the spectrum is preserved~\cite{Bla18,Zel20} in systems with either broken and unbroken $PT$-symmetry, extending the conventional systems with $PT$-symmetry and real spectrum~\cite{Zno00,Bag01,Cor15,Con19}. 

On the other hand, the Painlev\'e transcendents form a family of ordinary nonlinear equations that have become a topic under intensive research within both the mathematician and physicists communities. Their solutions cover a wide range of mathematical models with well-defined monodromy and also emerge naturally in several physical problems in the classical and quantum regimes. The Painlev\'e transcendents are characterized by a family of six nonlinear equations $P_{\rm I}$-$P_{\rm VI}$ defined in terms of complex parameters, whose solutions, in general, cannot be expressed in terms of classical functions~\cite{Gro02,Mar06}. Nevertheless, for some specific values of the parameters, a seed function can be used to generate a complete hierarchy of solutions through the B\"acklund transformation~\cite{Bas95}, which can be thought as a nonlinear counterpart of the recurrence relations. In particular, the fourth Painlev\'e equation can be taken into a Riccati equation with the appropriate choice of the parameters~\cite{Cla08}, where it solves a ``simpler'' nonlinear equation instead. Moreover, the fourth Painlev\'e equation has also brought new results in the trend of orthogonal polynomials, where new families were discovered through the hierarchies of rational solutions in terms of the generalized Okamoto, generalized Hermite and Yablonskii-Vorob'ev polynomials~\cite{Cla03}. The Painlev\'e transcendents have also found interesting applications in the study of physical models in nonlinear optics~\cite{Flo90}, quantum gravity~\cite{Fok91}, and SUSYQM~\cite{Mar09,Mar09a,Ber11,Ber13,Ber16}.

In particular, the fourth Painlev\'e equation arises quite naturally in third-order shape-invariant SUSYQM~\cite{Gen83,And00}, where the parameters of the Painlev\'e transcendent define the eigenvalues of the new Hamiltonians and the respective intertwining operators serve at the same time as ladder operators. Strinkingly, the so-constructed intertwining operators are not in general factorizable in terms of first-order operators. Thus, the results obtained in this way generalize those of~\cite{Suk97}. It is worth to notice that higher-order ladder operators have been also studied, in a different way, in the context of supersymmetric (SUSY) partners for the stationary oscillator in both the Hermitian~\cite{Fer99,Fer07} and non-Hermitian regimes~\cite{Ros18}. The Hermitian construction is particularly interesting from the mathematical point of view, for it has been exploited to construct and study new families of \textit{exceptional orthogonal polynomials}~\cite{Kui15,Gom20}, that is, a set of orthogonal polynomials with some degrees absent from the polynomial sequence. Such exceptional polynomials find applications in the construction and study of minimal surfaces~\cite{Cha20}.

In this manuscript, we develop a thorough study of quantum models that satisfy a third-order shape-invariant condition so that the corresponding potential becomes a rational extension of the harmonic oscillator written in terms of the Okamoto polynomials. The latter is achieved through the use of the fourth Painlev\'e equation, from which the solutions of the ``$-2x/3$'' hierarchy of rational solutions is exploited. Although some works have been published in the past for a wide range of solutions to the fourth Painkev\'e equation, the rational case was studied only for some specific values of the paramaters $\alpha$ and $\beta$ that define the fourth Painlev\'e transcendent~\cite{Mar09a,Mar16,Zel20c}. Thus, we provide the most general construction in terms of the ``$-2x/3$'' hierarchy. Our construction leads to three sequences of solutions, where the three eigenfunctions annihilated by the annihilation operator (zero-modes) are written in terms of the generalized Okamoto polynomials. The rest of the eigenfunctions (higher-modes) are computed after identifying a three-term recurrence relation for each sequence. Such higher-modes are determined iteratively after fixing the corresponding zero-mode as the initial condition. Interestingly, the latter leads to a relationship between the higher-modes and the family exceptional Hermite polynomials $H_{\lambda^{2},n}$ defined by the particular double partition $\lambda^{2}\equiv\lambda_{k}^{2}=(1,1,2,2,\cdots,k,k)$. This provides us with a set of three different three-term recurrence relations that generate the set of exceptional Hermite polynomials $\{H_{\lambda_{k},n}\}_{n=0}^{\infty}$. In previous works~\cite{Gom14,Gom16,Oda13}, several constructions for the exceptional Hermite polynomials have been identified through higher-order recurrence relations (of order higher than three), even for the double partition $\lambda^{2}_{k}$. Thus, our approach brings a new simple construction for such exceptional polynomials in the form of three-term recurrence relations, a property akin to classical orthogonal polynomials.

The manuscript is structured as follows. In Sec.~\ref{sec:thirdSI}, we briefly discuss the construction of third-order shape invariant Hamiltonians and their relation to the fourth Painlev\'e equation, with a particular emphasis to the ``$-2x/3$'' hierarchy of rational solutions. For such a case, the eigenvalues and zero-modes are determined in Sec.~\ref{sec:spectrum} in general. In Sec.~\ref{sec:highermodes}, we introduce an algorithm to construct the higher-modes from the zero-modes through a set of three-term recurrence relations. In this form, the complete spectral information can be determined. Interestingly, in Sec.~\ref{sec:SUSY}, we establish a relation between the third-order shape-invariant potential previously constructed and a higher-order Darboux-Crum transformation, which leads to a Wronksian representation of both the potential and eigenfunctions. The latter allows defining a three-term recurrence relation for a particular family of exceptional orthogonal Hermite polynomials. In App.~\ref{sec:Backlund}, we summarize some useful B\"acklund transformations for the fourth Painlev\'e transcendent that lead to the identities used throughout the manuscript. In App.~\ref{sec:APPB}, with aid of the Wronskian representation, we compute the explicit action of the creation operator on the eigenfunctions. Lastly, in App.~\ref{sec:APPC}, we summarize some basic properties of the Darboux-Crum (higher-order SUSY) transformations.
%---------------------------------------------------

%---------------------------------------------------
\section{Third-order shape-invariant systems}
\label{sec:thirdSI}
For completeness, in this section, we briefly discuss the construction of families of stationary Hamiltonians satisfying the third-order shape-invariant condition. Those models were initially identified in~\cite{And95}, recently studied in~\cite{Mar09a,Can00,Mar16} for the stationary case, and in~\cite{Zel20c} for the generalized time-dependent regime. In such cases, the construction of the corresponding potentials has been handled with generality. Nevertheless, the complete set of eigenfunctions has not been identified explicitly, and only the zero-modes have been presented in some particular cases. Thus, we focus on the general construction of rationally extended potentials determined in terms of the Okamoto polynomials.

Let us consider the eigenvalue equation related to an unknown Hamiltonian $H$ of the form
\begin{equation}
H\phi_{n}=E_{n}\phi_{n} \, , \quad H:=-\frac{d^{2}}{dx^{2}}+V(x) \, ,
\label{H}
\end{equation}
with $\phi_{n}(x)$ the eigenfunctions and $E_{n}$ the respective eigenvalues. The real-valued potential $V(x)$ is unknown and determined from a shape-invariant condition
\begin{equation}
HA^{\dagger}=A^{\dagger}(H+2\lambda) \, , \quad HA=A(H-2\lambda) \, , \quad \lambda>0 \, .
\label{shapeH}
\end{equation}
In the latter, $A$ and $A^{\dagger}$  are known as the \textit{intertwining operators}. In this particular case, we are interested in the third-order shape-invariant condition~\cite{Gen83} so that the intertwining operators, in coordinate representation, are represented as
\begin{align}
&A:=\frac{d^{3}}{dx^{3}}+A_{2}(x)\frac{d^{2}}{dx^{2}}+A_{1}(x)\frac{d}{dx}+A_{0}(x) \, , 
\label{A}\\
&A^{\dagger}:= \frac{d^{3}}{dx^{3}}+A_{2}(x)\frac{d^{2}}{dx^{2}}+\left[2A'_{2}(x)-A_{1}(x)\right]\frac{d}{dx}+\left[A_{0}(x)-A_{1}'(x)+A_{2}''(x)\right] \, , 
\label{AD}
\end{align}
with $A_{i}(x)$ real-valued functions, for $i=0,1,2$, to be determined from Eq.~\eqref{shapeH}, along with $f'(x)\equiv df/dx$ and $f''(x)\equiv d^{2}f/dx^{2}$.

Also, it is worth to remark that Eq.~\eqref{shapeH} also implies that $A$ and $A^{\dagger}$ are annihilation and creation operators, respectively, where the parameter $\lambda$ dictates the growth rate of the energy eigenvalues $E_{n}$, which increase or decrease by $2\lambda$ units. Such a parameter can also be absorbed after a suitable reparametrization of the spatial coordinate. Thus, without loss of generality, we consider $\lambda=1$.

Although the intertwining operators $A$ and $A^{\dagger}$ given in Eqs.~\eqref{A}-\eqref{AD} have the most general form, we consider a convenient factorization so that the Painlev\'e transcendent emerges naturally. The latter was first introduced by Cannata \textit{et al.}~\cite{Can00}, and it is usually known as an \textit{irreducible factorization} of the form
\begin{equation}
A=M^{\dagger}Q \, ,\quad A^{\dagger}=Q^{\dagger}M \, ,
\label{Afact1}
\end{equation}
where the sets $\{Q,Q^{\dagger}\}$ and $\{M,M^{\dagger} \}$  contain first-order and second-order operators, respectively, given by
\begin{align}
&M^{\dagger}:=\frac{d}{dx^{2}}-2G(x)\frac{d}{dx}+B(x) \, , \quad &&M=\frac{d}{dx^{2}}+2G(x)\frac{d}{dx}+B(x)+2G'(x) \, , 
\label{opM}\\
&Q^{\dagger}:=\frac{d}{dx}+W(x) \, ,\quad &&Q=-\frac{d}{dx}+W(x) \, ,
\label{opQ}
\end{align}
with $W(x)$, $B(x)$, and $G(x)$ real-valued functions to be determined. In  analogy to the relations given in Eq.~\eqref{shapeH}, the new operators in Eqs.~\eqref{opM}-\eqref{opQ} define an alternative set of intertwining relations that are not of the shape-invariant type. To see the latter, let us consider an auxiliary Hamiltonian of the form
\begin{equation}
H_{1}:=-\frac{d^{2}}{dx^{2}}+V_{1}(x) \, , \quad H_{1}\varphi_{n}^{(1)}=E_{n}^{(1)}\varphi_{n}^{(1)} \, , 
\end{equation}
with $V_{1}(x)$ the corresponding potential, together with the eigenfunctions $\varphi_{n}^{(1)}$ and eigenvalues $E_{n}^{(1)}$. We thus impose the following intertwining relations:
\begin{align}
& HQ^{\dagger}=Q^{\dagger}(H_{1}+2) \, , && HQ=Q(H_{1}-2) \, ,
\label{intertQ} \\
& HM^{\dagger}=M^{\dagger}H_{1} \, , &&HM=MH_{1} \, ,
\label{intertM}
\end{align}
such that their combined action bring us back to the original shape-invariant condition~\eqref{shapeH}, with $\lambda=1$. 

Now, substituting~\eqref{opM}-\eqref{opQ} into~\eqref{intertQ}-\eqref{intertM}, we obtain a set of differential equations involving the potentials $V(x)$ and $V_{1}(x)$ in terms of the unknown functions $B$, $G$, and $W$. The straightforward calculations show that
\begin{align}
&V(x)=V_{1}(x)+2W'+2 \, , &&V_{1}(x)=-W'+W^{2}-2 \, , 
\label{VV1}\\
&V(x)=V_{1}(x)-4G' \, , &&V_{1}(x)=2G^{2}+G'-B+\gamma \, ,
\label{VV11}
\end{align}
\begin{equation}
B=G^{2}-G'-\frac{G''}{2G}+\frac{(G')^{2}}{4G^{2}}+\frac{d}{4G^{2}} \, ,
\label{BG}
\end{equation}
where $d,\gamma\in\mathbb{R}$ are integration constants. After combining~\eqref{VV1}-\eqref{VV11} we obtain the two additional relations
\begin{equation}
W=-(2G+x) \, , \quad B=W'-W^{2}+G'+2G^2+\gamma+2 \, ,
\label{Wg}
\end{equation}
which combined with~\eqref{BG} leads to
\begin{equation}
G''=\frac{(G')^{2}}{2G}+6G^{3}+8 x G^{2}+2[x^{2}-(\gamma+1)] G+\frac{d}{2G} \, .
\label{eqg}
\end{equation}
In this form, the fourth Painlev\'e~\cite{Bas95,Cla03,Olv10} transcendent
\begin{equation}
w''=\frac{(w')^{2}}{2w}+\frac{3}{2}w^{3}+4xw^{2}+2(x^2-\alpha)w+\frac{\beta}{w} \, ,
\label{eqw}
\end{equation}
is recovered from~\eqref{eqg} through the reparametrization 
\begin{equation}
G(x)=\frac{w(x)}{2} \, , \quad \alpha=\gamma+1 \, , \quad \beta=2d \, .
\label{Gw}
\end{equation}
Although we are only interested in the construction of $V(x)$, we can also provide an explicit expression for $V_{1}(x)$. From~\eqref{VV1} and~\eqref{Wg}, we get the potentials in terms of the solutions $w(x)$ to the fourth Painlev\'e equation~\eqref{eqw} as
\begin{align}
&V(x)=x^{2}-\left(w'-2xw-w^{2}\right)-1 \, , 
\label{V}\\
&V_{1}(x)= x^{2}+(w'+2xw+w^{2})-1 \, .
\label{V1}
\end{align}
The latter reveals that $V(x)$ and $V_{1}(x)$ are \textit{regular functions}, that is, free of singularities, as long as $w(x)$ is a regular function as well. From now on, we focus on the properties of $w(x)$ that guarantee the desired regularity.

The fourth Painlev\'e transcendent~\eqref{eqw} has been extensively studied in the literature~\cite{Bas95,Cla03,Gro02}. Following some previous reports on the construction of third-order shape invariant quantum models~\cite{Mar09,Zel20c}, we particularly focus on the rational solutions of Eq.~\eqref{eqw}. Such a case has been studied for only some specific values of the parameters $\alpha$ and $\beta$, where a connection with a two-step Darboux-Crum transformation was found in~\cite{Mar09}. So far, a generalization for arbitrary values of the parameters is still absent. The complete set of rational solutions to the fourth Painlev\'e equation are classified into three hierarchies~\cite{Bas95,Cla03}, namely the ``$-1/x$ hierarchy'', ``$-2x$ hierarchy'', and the ``$-2x/3$ hierarchy''. The first two hierarchies contain solutions in terms of the \textit{generalized Hermite polynomials}~\cite{Cla03}, which have been already considered in previous works to obtain regular potentials for some particular rational extensions of the harmonic oscillator~\cite{Mar09a,Mar16}. 

In this work, we focus on the ``$-2x/3$'' hierarchy of rational solutions that lead to regular physical models. Rational solutions in such hierarchy are divided into three different types of solutions of the form~\cite{Bas95,Mar06}
\begin{align}
&w^{[1]}_{m,n}(x)=-\frac{2x}{3}+\frac{d}{dx}\ln\frac{Q_{m+1,n}}{Q_{m,n}} \, , \quad &&\alpha=2m+n \, , \quad &&\beta=-2\left(n-\frac{1}{3}\right)^{2} \, ,
\label{w1}\\
&w^{[2]}_{m,n}(x)=-\frac{2x}{3}+\frac{d}{dx}\ln\frac{Q_{m,n}}{Q_{m,n+1}} \, , \quad &&\alpha=-m-2n \, , \quad &&\beta=-2\left(m-\frac{1}{3}\right)^{2} \, ,
\label{w2}\\
&w^{[3]}_{m,n}(x)=-\frac{2x}{3}+\frac{d}{dx}\ln\frac{Q_{m,n+1}}{Q_{m+1,n}} \, , \quad &&\alpha=n-m \, , \quad &&\beta=-2\left(m+n+\frac{1}{3}\right)^{2} \, ,
\label{w3}
\end{align}
where $Q_{m,n}\equiv Q_{m,n}(x)$ stands for the \textit{generalized Okamoto polynomials}\cite{Oka86}, computed from the nonlinear recurrence relations
\begin{align}
Q_{m+1,n}Q_{m-1,n}=\frac{9}{2}\left[Q_{m,n}Q_{m,n}''-(Q_{m,n}')^{2}\right]+\left[2x^{2}+3(2m+n-1)\right]Q_{m,n}^{2} \, , 
\label{recurrenceQ1}\\
Q_{m,n+1}Q_{m,n-1}=\frac{9}{2}\left[Q_{m,n}Q_{m,n}''-(Q_{m,n}')^{2}\right]+\left[2x^{2}+3(1-m-2n)\right]Q_{m,n}^{2} \, ,
\label{recurrenceQ2}
\end{align}
for $m,n=0,1,\cdots$, together with the initial conditions $Q_{0,0}=Q_{1,0}=Q_{0,1}=1$ and $Q_{1,1}=\sqrt{2}\,x$. Moreover, with the aid of the B\"acklund transformations, the rational solutions~\eqref{w1}-\eqref{w3} can be cast into the new equivalent forms (see App.~\ref{sec:Backlund} for details)
\begin{align}
&w^{[1]}_{m,n}(x)=-\frac{\sqrt{2}}{3} \, \frac{Q_{m+1,n-1}Q_{m,n+1}}{Q_{m,n}Q_{m+1,n}} \, , 
\label{w1r}\\
&w^{[2]}_{m,n}(x)=-\frac{\sqrt{2}}{3} \, \frac{Q_{m+1,n}Q_{m-1,n+1}}{Q_{m,n+1}Q_{m,n}} \, ,
\label{w2r}\\
&w^{[3]}_{m,n}(x)=-\frac{\sqrt{2}}{3} \, \frac{Q_{m+1,n+1}Q_{m,n}}{Q_{m+1,n}Q_{m,n+1}} \, .
\label{w3r}
\end{align}

%-------->
\begin{table}
\centering
\footnotesize{\begin{tabular}{|c|p{15cm}|}
\hline
\rowcolor{gray} $\textcolor{white}{k}$ & $\textcolor{white}{Q_{k}(x)}$ \\
\hline
 0 & $1$ \\
 1 & $1$ \\
 2 & $2 x^2+3$ \\
 3 & $8 x^6+60 x^4+90 x^2+135$ \\
 4 & $64 x^{12}+1344 x^{10}+9360 x^8+30240 x^6+56700 x^4+170100 x^2+127575$ \\
 5 & $1024 x^{20}+46080 x^{18}+817920 x^{16}+7603200 x^{14}+41731200 x^{12}+155675520 x^{10}+493970400 x^8+1886068800 x^6+5304568500 x^4+5304568500 x^2+3978426375$\\
\hline
\end{tabular}}
\caption{Conventional Okamoto polynomials $Q_{k}(x)$ computed from~\eqref{recurrenceQ1} for $m=k$, $n=0$, and the initial conditions $Q_{0}=Q_{1}=1$.}
\label{tabQk}
\end{table}
%-------->

Among the generalized Okamoto polynomials, only the Okamoto polynomials $Q_{k}\equiv Q_{k,0}$ have zeros distributed outside of the real line. For details, see the zeros distribution in~\cite{Cla03} and the analysis provided in Sec.~\ref{subsec:Wzeros}. Thus, from Eqs.~\eqref{w1}-\eqref{w3}, regular rational solutions to the fourth Painlev\'e equation $\tilde{w}_{k}(x)$ are obtained only if 
\begin{equation}
\tilde{w}_{k}(x)\equiv w(x) \equiv w^{[1]}_{k,0}(x)=-\frac{2x}{3}+\frac{d}{dx}\ln\frac{Q_{k+1}}{Q_{k}} \, , \quad \alpha\equiv\tilde{\alpha}=2k \, , \quad \beta\equiv\tilde{\beta}=-2/9 \, ,
\label{wk}
\end{equation}
for $ k=0,1,\cdots$. Interestingly, with the aid of the B\"acklund and Schlesinger transformations~\cite{Cla08} (see App.~\ref{sec:Backlund} for details), along with the identities~\eqref{w1r}-\eqref{w3r}, the potential $V(x)$ defined in~\eqref{V} takes the form
\begin{equation}
V^{(k)}(x)\equiv V(x)=x^{2}-\left(\tilde{w}_{k}'-2x\tilde{w}_{k}-\tilde{w}_{k}^{2}\right)-1=x^{2}-\frac{4}{9}\frac{Q_{k+2}Q_{k}}{Q_{k+1}^{2}}+4k+1 \, ,
\label{potVk}
\end{equation}
and the recurrence relation~\eqref{recurrenceQ1} leads to the equivalently expression
\begin{equation}
V^{(k)}(x)\equiv \frac{x^{2}}{9}-2\frac{d^{2}}{dx^{2}}\ln Q_{k+1}+\frac{4k}{3}-\frac{1}{3} \, .
\label{potVk2}
\end{equation}
Eqs.~\eqref{potVk}-\eqref{potVk2} are the most general forms of the rationally extended potentials generated from the ``$-2x/3$'' hierarchy. In particular, using the Okamoto polynomials presented in Tab.~\ref{tabQk}, we recover the potential $V^{(k=1)}(x)$ reported in~\cite{Mar09}, and $V^{(k=2)}(x)$ reported in~\cite{Zel20c}. Besides the generality of the potential in~\eqref{potVk}, we can explicitly determine the eigenfunctions $\phi_{n}$ and eigenvalues $E_{n}$ in~\eqref{H}. Details are presented in the following sections.
%---------------------------------------------------

%---------------------------------------------------
\section{Mapping operators: zero-modes and eigenvalues}
\label{sec:spectrum}
As discussed in the previous section, the third-order intertwining operators $A$ and $A^{\dagger}$ have been factorized in terms of the second-order and first-order operators introduced in~\eqref{Afact1} to explicitly show the appearance of the fourth Painlev\'e equation and its relation to the potential $V(x)$. Nevertheless, to study the spectral information of $H$, it is more convenient and simple if we further decompose the second-order operators as $M^{\dagger}=M_{1}^{\dagger}M_{2}^{\dagger}$ and $M=M_{2}M_{1}$, where the first-order operators $M_{1,2}$ and $M_{1,2}^{\dagger}$ are of the form
\begin{align}
&M_{1}:=-\frac{d}{dx}+W_{1}(x) \, , \quad &&M_{2}:=-\frac{d}{dx}+W_{2}(x) \, , 
\label{M1M2}\\
&M^{\dagger}_{1}:=\frac{d}{dx}+W_{1}(x) \, , \quad &&M^{\dagger}_{2}:=\frac{d}{dx}+W_{2}(x) \, ,
\label{M1M2D}
\end{align}
where the real-valued functions $W_{1,2}(x)$ are computed from the factorization $M^{\dagger}=M_{1}^{\dagger}M_{2}^{\dagger}$ after comparing term by term with~\eqref{opM}, and using~\eqref{Wg}. The particular factorization discussed above corresponds to the \textit{reducible case} discussed in~\cite{And95,And00}. The straightforward calculations lead to~\cite{Mar16,Zel20c}
\begin{equation}
W^{\pm}_{1}(x)\equiv W_{1}(x)=-G+\frac{G'\pm\sqrt{-d}}{2G} \, , \quad W^{\pm}_{2}(x)\equiv W_{2}(x)=-G-\frac{G'\pm\sqrt{-d}}{2G} \, ,
\end{equation}
which, with the aid of~\eqref{Gw}, take the more convenient form
\begin{equation}
W^{\pm}_{1}(x)=\frac{\tilde{\mathcal{F}}_{k}^{-}\pm\sqrt{-2\tilde{\beta}}}{2\tilde{w}}+x \, , \quad W^{\pm}_{2}(x)=-\left(\frac{\tilde{\mathcal{F}}_{k}^{+}\pm\sqrt{-2\tilde{\beta}}}{2\tilde{w}}\right)+x \, , 
\label{W1W2-1}
\end{equation}	
where $\tilde{\mathcal{F}}_{k}^{\pm}=\tilde{w}'_{k}\pm(2x\tilde{w}_{k}+\tilde{w}_{k}^{2})$, with $\tilde{w}_{k}$ given in Eq.~\eqref{wk}. In this form, the B\"acklund transformations introduced in~\eqref{w1pm}, combined with $w_{0}=\tilde{w}_{k}$, allows us to simplify~\eqref{W1W2-1} into
\begin{align}
& W_{1}^{+}=\frac{x}{3}+\frac{d}{dx}\ln\frac{Q_{k,1}}{Q_{k+1}} \, , && W_{1}^{-}=\frac{x}{3}+\frac{d}{dx}\ln\frac{Q_{k+1,-1}}{Q_{k+1}} \, , \\
& W_{2}^{+}=\frac{x}{3}+\frac{d}{dx}\ln\frac{Q_{k}}{Q_{k,1}} \, , && W_{2}^{-}=\frac{x}{3}+\frac{d}{dx}\ln\frac{Q_{k}}{Q_{k+1,-1}} \, .
\label{W1W2pm}
\end{align}

Notice that $W_{1}^{-}$ and $W_{2}^{-}$ depend on generalized Okamoto polynomials $Q_{m,-1}$ that contain a negative index, which are determined iteratively from the recurrence relation~\eqref{recurrenceQ2} with $n=0$. For instance, $Q_{0,-1}$ is computed from $Q_{0,0}=Q_{0,1}=1$, $Q_{1,-1}$ from $Q_{1,1}$ and $Q_{1,0}$, $Q_{2,-1}$ from $Q_{2,1}$ and $Q_{2,0}$, and so on.

\subsection{Mapping operators}
\label{subsec:map-op}

%--------------------->
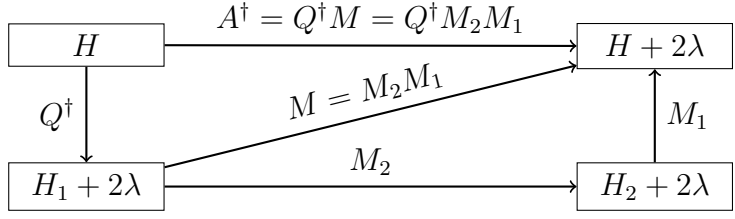
\begin{figure}
\centering
\begin{tikzpicture}
\tikzstyle{line} = [draw, thick, -latex’,shorten >=2pt];
\matrix (m) [matrix of nodes,row sep=1.5em,column sep=3em,minimum width=5em]
{
\node(m21) [rectangle,draw] {$H$};	& \node(XX) [minimum width=30mm] {};	    & \node(m22) [rectangle,draw] {$H+2\lambda$}; \\
     	    &    		&   				  & \\
\node(m32) [rectangle,draw] {$H_{1}+2\lambda$}; & & \node(m34) [rectangle,draw] {$H_{2}+2\lambda$}; 
\\};
\path[->,black,thick] (m21) edge node[anchor=center, left] {$Q^{\dagger}$} (m32);
\path[->,black,thick] (m21) edge node[anchor=center, above] {$A^{\dagger}=Q^{\dagger}M=Q^{\dagger}M_{2}M_{1}$}  (m22);
\path[->,black,thick] (m32) edge node[sloped, anchor=center, above] {$M=M_{2}M_{1}$} (m22);
\path[->,black,thick] (m32) edge node[anchor=center, above] {$M_{2}$} (m34);
\path[->,black,thick] (m34) edge node[anchor=center, right] {$M_{1}$} (m22);
\end{tikzpicture}
\caption{\footnotesize{Third-order shape-invariant Hamiltonian $H$. The arrows indicate the intertwining relation among the different Hamiltonians $H$, $H_{1}$, and $H_{2}$. For instance, the arrow on top indicates $HA^{\dagger}=A^{\dagger}(H+2\lambda)$, as given in~\eqref{shapeH}. The direction of arrows is inverted by using the adjoint relations.}}
\label{diagram}
\end{figure}
%------------------------>

In Sec.~\ref{sec:thirdSI}, it was shown that the factorization operators $M$, $M^{\dagger}$, $Q$, and $Q^{\dagger}$ define intertwining relations between $H$ and the auxiliary Hamiltonian $H_{1}$. Similar relations hold for the factorized operators introduced in~\eqref{M1M2}-\eqref{M1M2D}. In analogy to the procedure followed in the previous section, we introduce another auxiliary Hamiltonian $H_{2}$, together with the corresponding eigenvalue equation,
\begin{equation}
H_{2}:=-\frac{d^{2}}{dx^{2}}+V_{2}(x) \, , \quad H_{2}\varphi_{n}^{(2)}=E_{n}^{(2)}\varphi_{n}^{(2)} \, ,
\label{H2}
\end{equation}
with $\varphi_{n}^{(2)}$ and $E_{n}^{(2)}$ the eigenfunctions and eigenvalues, respectively, so that $H_{2}$ intertwines with $H$ and $H_{1}$ through $M_{1}$, $M_{1}^{\dagger}$, $M_{2}$, and $M_{2}^{\dagger}$ as
\begin{align}
& H_{1}M_{2}=M_{2}H_{2} \, , && H_{2}M_{2}^{\dagger}=M_{2}^{\dagger}H_{1} \, ,
\label{intertH1M2}\\
& H_{2}M_{1}=M_{1}H \, , && HM_{1}^{\dagger}=M_{1}^{\dagger}H_{2} \, .
\label{intertH2M1}
\end{align}
Clearly, the combined action of~\eqref{intertH1M2}-\eqref{intertH2M1} brings us back to the initial intertwining relation introduced in~\eqref{intertM}. Thus, for clarity, the action of the intertwining relations constructed so far is depicted and summarized in Fig.~\ref{diagram}, where it is also evident that $A^{\dagger}$ and $A$ are ladder operators in $\mathcal{H}$. Furthermore, Fig.~\ref{diagram} reveals that the elements of the set $\{ M_{1}, M_{1}^{\dagger}, M_{2}, M_{2}^{\dagger}, Q, Q^{\dagger} \}$ define mapping operators among the vector spaces generated by the eigenfunctions of $H$ and the auxiliary Hamiltonians $H_{1}$ and $H_{2}$. We thus get
\begin{equation}
\begin{alignedat}{3}
& M_{2}^{\dagger}:\mathcal{H}_{1}\rightarrow\mathcal{H}_{2} \, , \quad && M_{2}:\mathcal{H}_{2}\rightarrow\mathcal{H}_{1} \, , \\
& M_{1}^{\dagger}:\mathcal{H}_{2}\rightarrow\mathcal{H}\, , \quad && M_{1}:\mathcal{H}\rightarrow\mathcal{H}_{2} \, , \\
& Q^{\dagger}:\mathcal{H}_{1}\rightarrow\mathcal{H}\, , \quad && Q:\mathcal{H}\rightarrow\mathcal{H}_{1} \, .
\end{alignedat}
\label{mappings}
\end{equation}
In the latter, the vector spaces are defined as
\begin{equation}
\mathcal{H}=\operatorname{Span}\{\phi_{n}\}_{n=0}^{\infty} \, , \quad \mathcal{H}_{1}=\operatorname{Span}\{\varphi_{n}^{(1)}\}_{n=0}^{\infty} \, , \quad \mathcal{H}_{2}=\operatorname{Span}\{\varphi_{n}^{(2)}\}_{n=0}^{\infty} \, .
\label{vec-spaces}
\end{equation}

Before concluding this section, it is worth noticing that the relations~\eqref{intertH1M2}-\eqref{intertH2M1} can be further exploited by recalling that $M_{1,2}$ and $M_{1,2}^{\dagger}$ are first-order operators, and thus can be exploited to get~\cite{Zel20c}
\begin{equation}
\begin{alignedat}{3}
& H=M_{1}^{\dagger}M_{1}+\epsilon_{1} \, , \quad && H_{2}=M_{1}M_{1}^{\dagger}+\epsilon_{1} \, , \\
& H_{2}=M^{\dagger}_{2}M_{2}+\epsilon_{2} \, , && H_{1}=M_{2}M_{2}^{\dagger}+\epsilon_{2} \, ,
\end{alignedat}
\label{facI1I2}
\end{equation}
for some real constants $\epsilon_{1}$ and $\epsilon_{2}$ determined after
substituting Eqs.~\eqref{M1M2}-\eqref{M1M2D} and Eqs.~\eqref{W1W2-1} into~\eqref{facI1I2}. After some calculations we obtain
\begin{equation}
\epsilon_{1}=\gamma-\sqrt{-d}=2k-\frac{4}{3} \, , \quad \epsilon_{2}=\gamma+\sqrt{-d}=2k-\frac{2}{3} \, .
\label{e1e2}
\end{equation}
%------------------------------------>

%------------------------------------>
\subsection{Zero-modes $\phi_{0;j}^{(k)}$}
\label{subsec:zeromode}
With the operations depicted in Fig.~\ref{diagram}, we determine the spectral information of $H$. Although the eigenvalues of the auxiliary Hamiltonians $H_{1}$ and $H_{2}$ can be obtained in a similar way, the latter is not relevant for the present work and thus will be omitted. Let us recall that $A=M_{1}^{\dagger}M_{2}^{\dagger}Q$ is an annihilation operator, which is also of third-order, and thus three different eigenfunctions are simultaneously annihilated by it, that is,
\begin{equation}
A\phi_{0;j}^{(k)}=M^{\dagger}Q\phi_{0;j}^{(k)}=M_{1}^{\dagger}M_{2}^{\dagger}Q\phi_{0;j}^{(k)}=0 \, , \quad j=1,2,3.
\label{kernelA}
\end{equation}
Henceforth, the eigenfunctions $\phi_{0;j}^{(k)}$ will be called \textit{zero-modes}. Those functions are computed from the mappings defined in~\eqref{mappings} along with the factorization of the Hamiltonians $H$, $H_{1}$, and $H_{2}$. From~\eqref{kernelA} we distinguish three different cases:

\begin{itemize}
\item $\phi_{0;1}^{(k)}$: The zero-mode $\phi_{0;1}^{(k)}\in\mathcal{H}$ is annihilated by $Q$, that is, $Q\phi_{0;1}=0$.
\item $\phi_{0;2}^{(k)}$: $Q$ does not annihilate the zero-mode, $Q\phi_{0;2}^{(k)}=f_{2}\not=0$, instead we have $M_{2}^{\dagger}f_{2}=0$. From~\eqref{facI1I2} it follows that a function annihilated by $M_{2}^{\dagger}$ is a zero-mode of $H_{1}$, say $f_{2}\equiv\varphi^{(1)}_{0}\in\mathcal{H}_{1}$, which can be mapped into the unnormalized zero-mode $\phi_{0;2}\in\mathcal{H}$ through the action $Q^{\dagger}\varphi_{0}^{(1)}$. 
\item $\phi_{0;3}$: Neither $Q$ nor $M_{2}^{\dagger}$ annihilate the zero-mode, $M_{2}^{\dagger}Q\phi_{0;3}=f_{3}\not=0$, instead we have $M_{1}^{\dagger}f_{3}=0$. From~\eqref{facI1I2} we conclude that a function annihilated by $M_{1}^{\dagger}$ is a zero-mode of $H_{2}$, say $f_{3}\equiv\varphi_{0}^{(2)}\in\mathcal{H}_{2}$. Thus, the respective unnormalized zero-mode $\phi_{0;3}\in\mathcal{H}$ is determined with the mapping $Q^{\dagger}M_{2}\varphi_{0}^{(2)}$.
\end{itemize}

%---------->
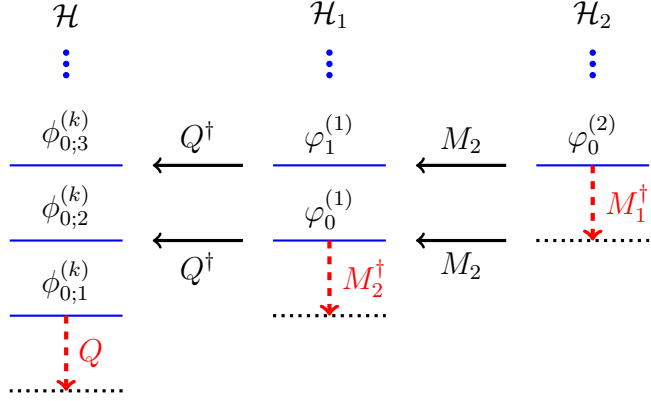
\begin{figure}
\centering
\begin{tikzpicture}
\draw (0.75,3) node {$\mathcal{H}$};
\draw (4.25,3) node {$\mathcal{H}_{1}$};
\draw (7.75,3) node {$\mathcal{H}_{2}$};
\filldraw[blue] (0.75,2.2) circle (1pt);
\filldraw[blue] (0.75,2.35) circle (1pt);
\filldraw[blue] (0.75,2.5) circle (1pt);
\filldraw[blue] (4.25,2.2) circle (1pt);
\filldraw[blue] (4.25,2.35) circle (1pt);
\filldraw[blue] (4.25,2.5) circle (1pt);
\filldraw[blue] (7.75,2.2) circle (1pt);
\filldraw[blue] (7.75,2.35) circle (1pt);
\filldraw[blue] (7.75,2.5) circle (1pt);
% Spectrum I1
\draw[blue,thick] (1.5,1)--(0,1) node[midway,anchor=south,text=black] {$\phi_{0;3}^{(k)}$};
\draw[blue,thick] (1.5,0)--(0,0) node[midway,anchor=south,text=black] {$\phi_{0;2}^{(k)}$};
\draw[blue,thick] (1.5,-1)--(0,-1) node[midway,anchor=south,text=black] {$\phi_{0;1}^{(k)}$};
\draw[black,very thick,dotted] (1.5,-2)--(0,-2);
\draw[red,ultra thick,dashed,->] (0.75,-1)--(0.75,-2) node[midway,anchor=west,text=red] {$Q$};
% Spectrum I2
\draw[blue,thick] (3.5,1)--(5,1) node[midway,anchor=south,text=black] {$\varphi_{1}^{(1)}$};
\draw[blue,thick] (3.5,0)--(5,0) node[midway,anchor=south,text=black] {$\varphi_{0}^{(1)}$};
\draw[black,very thick,dotted] (3.5,-1)--(5,-1);
\draw[red,ultra thick,dashed,->] (4.25,0)--(4.25,-1) node[midway,anchor=west,text=red] {$M_{2}^{\dagger}$};
% Spectrum J
\draw[blue,thick] (7,1)--(8.5,1) node[midway,anchor=south,text=black] {$\varphi^{(2)}_{0}$};
\draw[black,very thick,dotted] (7,0)--(8.5,0);
\draw[red,ultra thick,dashed,->] (7.75,1)--(7.75,0) node[midway,anchor=west,text=red] {$M_{1}^{\dagger}$};
% Arrows
%\draw[very thick,->] (1.9,0.6)--(3.1,0.6) node[midway,anchor=south] {$Q$};
\draw[very thick,->] (3.1,0)--(1.9,0) node[midway,anchor=north] {$Q^{\dagger}$};
\draw[very thick,->] (3.1,1)--(1.9,1) node[midway,anchor=south] {$Q^{\dagger}$};
%\draw[very thick,->] (5.4,1.1)--(6.6,1.1) node[midway,anchor=south] {$M_{2}^{\dagger}$};
\draw[very thick,->] (6.6,1)--(5.4,1) node[midway,anchor=south] {$M_{2}$};
\draw[very thick,->] (6.6,0)--(5.4,0) node[midway,anchor=north] {$M_{2}$};
\end{tikzpicture}
\caption{\footnotesize{Mappings between the zero-mode eigenfunctions $\phi_{0;j}^{(k)}$, for $j=1,2,3$, and the eigenfunctions of the auxiliary Hamiltonians $H_{1}$ and $H_{2}$. The dotted-black lines denote the null-vector.}}
\label{F2}
\end{figure}
%---------->

\begin{figure}
\centering
\subfloat[][$V^{(k)}(x)$]{\includegraphics[width=0.3\textwidth]{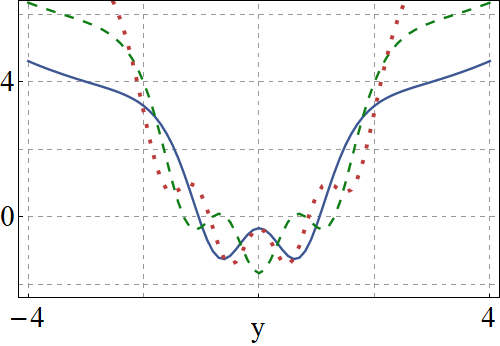}}
\hspace{2mm}
\subfloat[][$k=2$]{\includegraphics[width=0.31\textwidth]{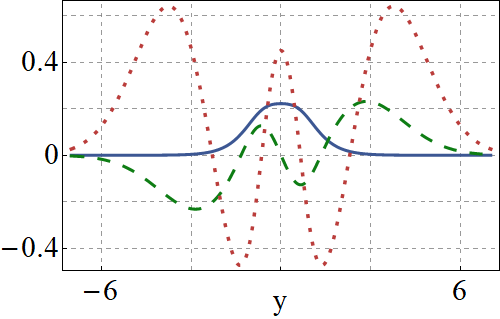}}
\hspace{2mm}
\subfloat[][$k=4$]{\includegraphics[width=0.31\textwidth]{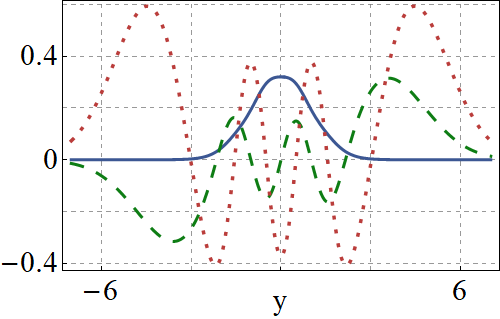}}
\caption{(a) Potential $V^{(k)}(x)$ given in \eqref{potVk} for $k=2$ (solid-blue), $k=3$ (dashed-green), and $k=4$ (dotted-red). The zero-modes $\phi_{0;1}$ (solid-blue), $\phi_{0;2}$ (dashed-green), and $\phi_{0;3}$ (dotted-red) are depicted for $k=2$ (b) and $k=4$ (c).}
\label{FigPotk}
\end{figure}

In Fig.~\ref{F2}, we depict a diagram summarizing the action of the mapping operators required to compute the zero-modes. The straightforward calculations lead to
\begin{equation}
\begin{aligned}
&\phi_{0;1}^{(k)}(x)=\mathcal{N}_{0;1}e^{\int dx W(x)} \, , \\
&\phi^{(k)\pm}_{0;2}(x)=\mathcal{N}_{0;2}\left(W-W^{\pm}_{2}\right)e^{-\int dx W^{\pm}_{2}(x)} \, , \\
&\phi^{(k)\pm}_{0;3}(x)=\mathcal{N}_{0;3}\left(\pm2\sqrt{-d}+(W-W^{\pm}_{2})(W^{\pm}_{1}+W^{\pm}_{2}) \right)e^{-\int dx W^{\pm}_{1}(x)} \, , 
\end{aligned}
\end{equation}
with the respective eigenvalues 
\begin{equation}
E_{0;1}=0 \, , \quad E^{\pm}_{0;2}=\gamma+2\mp\sqrt{-d} \, , \quad E^{\pm}_{0;3}=\gamma+2\pm\sqrt{-d} \, .
\label{EnergyZero}
\end{equation}
Notably, the zero-mode $\phi_{0;1}$ is easily determined in terms of $\tilde{w}_{k}$ through~\eqref{Wg} and
\begin{equation}
W=-(\tilde{w}_{k}+x)=-\left(\frac{x}{3}+\frac{d}{dx}\ln\frac{Q_{k+1}}{Q_{k}} \right) \, .
\end{equation}
We thus get
\begin{equation}
\phi_{0;1}^{(k)}(x)=\mathcal{N}_{0;1}\left(\frac{e^{-\frac{x^{2}}{6}}}{Q_{k+1}}\right)Q_{k}\, , \quad E_{0;1}=0 \, ,
\label{phi01}
\end{equation}
where the Okamoto polynomials $Q_{k}$ are presented in Table.~\ref{tabQk} for several values of $k$. It is worth recalling that $Q_{k}$ are nodeless functions for $x\in\mathbb{R}$. Thus, the zero-mode $\phi_{0;1}$ is both a regular and nodeless function. Thus, $\phi_{0;1}$ is the ground eigenfunction of $H$, with $E_{0;0}$ the lowest eigenvalue among all the sequences. Similarly, we compute the remaining zero-modes from~\eqref{W1W2pm}, leading to 
\begin{align}
&\phi_{0;2}^{(k)+}(x)=\mathcal{N}_{0;2}^{+}\left(\frac{e^{-\frac{x^{2}}{6}}}{Q_{k+1}} \right)Q_{k+1,1} \, , && \phi_{0;2}^{(k)-}(x)=\mathcal{N}_{0;2}^{-}\left(\frac{e^{-\frac{x^{2}}{6}}}{Q_{k+1}} \right)Q_{k+2,-1} \, , 
\label{phi02}\\
&\phi_{0;3}^{(k)+}(x)=\mathcal{N}_{0;3}^{+}\left(\frac{e^{-\frac{x^{2}}{6}}}{Q_{k+1}} \right)Q_{k+2,-1} \, , && \phi_{0;3}^{(k)-}(x)=\mathcal{N}_{0;3}^{-}\left(\frac{e^{-\frac{x^{2}}{6}}}{Q_{k+1}} \right)Q_{k+1,1} \, ,
\label{phi03}
\end{align}
where $\mathcal{N}_{0;2}^{\pm}$ and $\mathcal{N}_{0;3}^{\pm}$ stand for the normalization constants. From~\eqref{phi02}-\eqref{phi03}, it is clear that $\phi_{0;2}^{(k)\pm}=\phi_{0;3}^{(k)\mp}$ and $E_{0;2}^{\pm}=E_{0;3}^{\mp}$. Henceforth, we can freely fix, without loss of generality, the zero-modes and eigenvalues with the ``$+$'' superscript. To illustrate our results, we depict the behavior of the zero-modes $\phi_{0;j}^{(k)}$ and the corresponding potential $V^{(k)}(x)$ in Fig.~\ref{FigPotk} for several values of $k$.

%----------------->
%----------------->

%----------------->
\begin{table}
\centering
\footnotesize{\begin{tabular}{|c|p{15cm}|}
\hline
\rowcolor{gray} $\textcolor{white}{k}$ & $\textcolor{white}{Q_{k,1}(x)}$ \\
\hline
 0 & $1$ \\
 1 & $\sqrt{2} x$ \\
 2 & $4 x^4+12 x^2-9$ \\
 3 & $\sqrt{2}x(16 x^8+192 x^6+504 x^4-2835 ) $\\
 4 & $256 x^{16}+7680 x^{14}+80640 x^{12}+362880 x^{10}+453600 x^8-1905120 x^6-14288400 x^4-21432600 x^2+8037225$ \\
 5 & $\sqrt{2}x( 4096 x^{24}+245760 x^{22}+5990400 x^{20}+77414400 x^{18}+569721600 x^{16}+2246952960 x^{14}+1600300800 x^{12}-35663846400 x^{10}-275837562000 x^8-1103350248000 x^6-1737776640600 x^4+3258331201125)$ \\
 \hline
\end{tabular}}
\caption{Generalized Okamoto polynomials $Q_{m,n}$ computed from~\eqref{recurrenceQ1}-\eqref{recurrenceQ2} for $m=k$, $n=1$, and the initial conditions $Q_{0,1}=1$ and $Q_{1,1}=\sqrt{2}x$.}
\label{tabQk1}
\end{table}
%----------------->

%----------------->
\begin{table}
\centering
\footnotesize{\begin{tabular}{|c|p{15cm}|}
\hline
\rowcolor{gray} $\textcolor{white}{k}$ & $\textcolor{white}{Q_{k,-1}(x)}$ \\
\hline
 0 & $2 x^2+3$ \\
 1 & $\sqrt{2} x$ \\
 2 & $2 x^2-3$ \\
 3 & $\sqrt{2}x( 4 x^4-45) $\\
 4 & $32 x^{10}+240 x^8-720 x^6-5400 x^4-12150 x^2+6075 $\\
 5 & $\sqrt{2}x( 256 x^{16}+6144 x^{14}+34560 x^{12}-138240 x^{10}-2138400 x^8-8553600 x^6-22453200 x^4+63149625) $\\
% 6 & $8192 x^{26}+ 430080 x^{24}+ 8294400 x^{22} + 63866880 x^{20} -  79833600 x^{18} - 5340867840 x^{16} - 46702656000 x^{14}- 227406009600 x^{12} - 778003380000 x^{10} - 1367063082000 x^8 +  3501015210000 x^6 + 15754568445000 x^4 + 19693210556250 x^2 -5907963166875 $\\
\hline
\end{tabular}}
\caption{Generalized Okamoto polynomials $Q_{m,n}$ computed from~\eqref{recurrenceQ1}-\eqref{recurrenceQ2} for $m=k$, $n=-1$, and the initial conditions $Q_{0,-1}=2x^{2}+3$ and $Q_{1,-1}=\sqrt{2}x$.}
\label{tabQkm1}
\end{table}
%----------------->

So far, we have explicitly determined three regular zero-modes. Moreover, it is immediate to realize that each zero-mode converges asymptotically to zero, $\lim_{x\rightarrow\pm\infty}\phi_{0;j}\rightarrow 0$, for $j=1,2,3$. Thus the zero-modes become elements in the space of square-integrable functions, that is, $\phi_{0;j}\in L^{2}(dx;\mathbb{R})$ such that
\begin{equation}
(\phi_{0;j}^{(k)},\phi_{0;j}^{(k)}):=\int_{-\infty}^{\infty}\vert\phi_{0;j}^{(k)}(x)\vert^{2}<\infty \, , \quad j=1,2,3,
\end{equation}
with the inner product defined for any two eigenfunctions $\phi$ and $\tilde{\phi}$ as
\begin{equation}
(\tilde{\phi},\phi):=\int_{-\infty}^{\infty}dx \, [\tilde{\phi}(x)]^{*}\phi(x) \, .
\label{inn}
\end{equation}

In some cases, the eigenfunctions $\Psi_{0;j}^{(k)}$ annihilated by the creation operator $A^{\dagger}$ may lead to physical solutions as well. However, in this particular case it is straightforward to realize that any function $\Psi_{0;j}$, such that $A^{\dagger}\Psi_{0;j}^{(k)}=0$, diverges asymptotically at $x\rightarrow\pm\infty$ and thus $\Psi_{0;j}^{(k)}\not\in L^{2}(dx;\mathbb{R})$, for $j=1,2,3$. See~\cite{Zel20c} for additional details. Therefore, no physical eigenfunctions of $A^{\dagger}$ are extracted from the ``$-2x/3$'' hierarchy, and the spectrum is solely determined from the zero-modes~\eqref{phi01}-\eqref{phi03}.
%------------------------------------------------------->

%----------------------------------------------------
\section{Higher-modes $\phi_{n;j}^{(k)}$}
\label{sec:highermodes}
Given that $A^{\dagger}$ is a creation operator in $\mathcal{H}$, the \textit{higher-modes} $\phi_{n;j}$ are then computed through the iterated action on the zero-modes $\phi_{0;j}$, up to a normalization constant,  through $\phi_{n;j}\propto A^{\dagger \, n}\phi_{0;j}$, for $j=1,2,3$. From~\eqref{shapeH}, it follows that the eigenvalues are increased by two units after each iteration. In this form, the set of eigenvalues are determined straightforwardly from $E_{0;1}=0$, $E_{0;2}=2k+2/3$, and $E_{0;3}=2k+4/3$, leading to
\begin{equation}
E_{n;1}=2n \, , \quad E_{n;2}=2k+2n+2/3 \, , \quad E_{n;3}=2k+2n+4/3 \, .
\label{En}
\end{equation}
Notice that none of the eigenvalues overlaps; thus, the respective eigenfunctions should be, in general, all different. Since the creation operator $A^{\dagger}$ is represented by a third-order differential operator, any attempt to determine closed expressions for the higher-modes becomes a challenging task. 

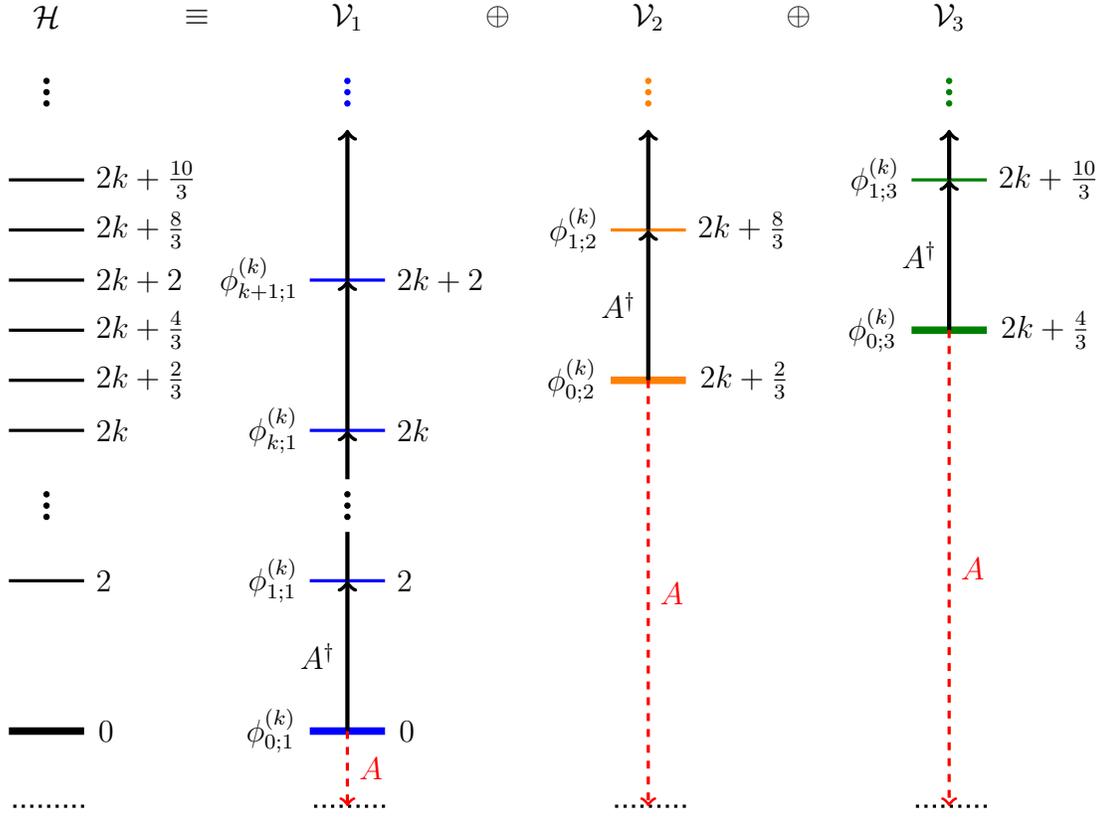
\begin{figure}[t]
\centering
\begin{tikzpicture}
\draw (1,9.5) node {$\mathcal{H}$};
\draw (5,9.5) node {$\mathcal{V}_{1}$};
\draw (9,9.5) node {$\mathcal{V}_{2}$};
\draw (13,9.5) node {$\mathcal{V}_{3}$};
\draw (3,9.5) node {$\equiv$};
\draw (7,9.5) node {$\oplus$};
\draw (11,9.5) node {$\oplus$};
% Black dots
\filldraw[black] (1,8.35) circle (1pt);
\filldraw[black] (1,8.5) circle (1pt);
\filldraw[black] (1,8.65) circle (1pt);
% Blue dots
\filldraw[blue] (5,8.35) circle (1pt);
\filldraw[blue] (5,8.5) circle (1pt);
\filldraw[blue] (5,8.65) circle (1pt);
% Orange dots
\filldraw[orange] (9,8.35) circle (1pt);
\filldraw[orange] (9,8.5) circle (1pt);
\filldraw[orange] (9,8.65) circle (1pt);
% Green dots
\filldraw[color=green!50!black] (13,8.35) circle (1pt);
\filldraw[color=green!50!black] (13,8.5) circle (1pt);
\filldraw[color=green!50!black] (13,8.65) circle (1pt);
% Null state
\draw[black,dotted,very thick] (1.5,-1)--(0.5,-1);
\draw[black,dotted,very thick] (5.5,-1)--(4.5,-1);
\draw[black,dotted,very thick] (9.5,-1)--(8.5,-1);
\draw[black,dotted,very thick] (13.5,-1)--(12.5,-1);
% Sequence 1
\draw[black,line width=1mm] (1.5,0)--(0.5,0) node[at start,anchor=west,text=black] {$0$} ;
\draw[black,very thick] (1.5,6/3)--(0.5,6/3) node[at start,anchor=west,text=black] {$2$} ;
\draw[black,very thick] (1.5,12/3)--(0.5,12/3) node[at start,anchor=west,text=black] {$2k$} ;
\draw[black,very thick] (1.5,18/3)--(0.5,18/3) node[at start,anchor=west,text=black] {$2k+2$} ;
\filldraw[black] (1,2.85) circle (1pt);
\filldraw[black] (1,3) circle (1pt);
\filldraw[black] (1,3.15) circle (1pt);
%Sequence 2
\draw[black,very thick] (1.5,14/3)--(0.5,14/3) node[at start,anchor=west,text=black] {$2k+\frac{2}{3}$};
\draw[black,very thick] (1.5,20/3)--(0.5,20/3) node[at start,anchor=west,text=black] {$2k+\frac{8}{3}$};
%Sequence 2
\draw[black,very thick] (1.5,16/3)--(0.5,16/3) node[at start,anchor=west,text=black] {$2k+\frac{4}{3}$};
\draw[black,very thick] (1.5,22/3)--(0.5,22/3) node[at start,anchor=west,text=black] {$2k+\frac{10}{3}$};
%\mathcal{V}_{1} spectrum
\draw[blue,line width=1mm] (4.5,0)--(5.5,0) node[anchor=west,text=black] {$0$} node[at start,anchor=east,text=black] {$\phi_{0;1}^{(k)}$};
\draw[blue,very thick] (4.5,2)--(5.5,2) node[anchor=west,text=black] {$2$} node[at start,anchor=east,text=black] {$\phi_{1;1}^{(k)}$};
\draw[blue,very thick] (4.5,4)--(5.5,4) node[anchor=west,text=black] {$2k$} node[at start,anchor=east,text=black] {$\phi_{k;1}^{(k)}$};
\draw[blue,very thick] (4.5,6)--(5.5,6) node[anchor=west,text=black] {$2k+2$} node[at start,anchor=east,text=black] {$\phi_{k+1;1}^{(k)}$};
\filldraw[black] (5,2.85) circle (1pt);
\filldraw[black] (5,3) circle (1pt);
\filldraw[black] (5,3.15) circle (1pt);
\draw[ultra thick,black,-] (5,2)--(5,2.65);
\draw[ultra thick,black,->] (5,3.35)--(5,4);
\draw[ultra thick,black,->] (5,0)--(5,2) node[midway,anchor=east] {$A^{\dagger}$};
\draw[ultra thick,black,->] (5,4)--(5,6);
\draw[ultra thick,black,->] (5,6)--(5,8);
\draw[red,dashed,very thick,->] (5,0)--(5,-1) node[midway,anchor=west] {$A$};
%\mathcal{V}_{2} spectrum
\draw[orange,line width=1mm] (8.5,14/3)--(9.5,14/3) node[anchor=west,text=black] {$2k+\frac{2}{3}$} node[at start,anchor=east,text=black] {$\phi_{0;2}^{(k)}$};
\draw[orange,very thick] (8.5,20/3)--(9.5,20/3) node[anchor=west,text=black] {$2k+\frac{8}{3}$} node[at start,anchor=east,text=black] {$\phi_{1;2}^{(k)}$};
\draw[ultra thick,black,->] (9,14/3)--(9,20/3) node[midway,anchor=east] {$A^{\dagger}$};
\draw[ultra thick,black,->] (9,20/3)--(9,8);
\draw[red,dashed,very thick,->] (9,14/3)--(9,-1) node[midway,anchor=west] {$A$};
%\mathcal{V}_{3} spectrum
\draw[color=green!50!black,line width=1mm] (12.5,16/3)--(13.5,16/3) node[anchor=west,text=black] {$2k+\frac{4}{3}$} node[at start,anchor=east,text=black] {$\phi_{0;3}^{(k)}$};
\draw[color=green!50!black,very thick] (12.5,22/3)--(13.5,22/3) node[anchor=west,text=black] {$2k+\frac{10}{3}$} node[at start,anchor=east,text=black] {$\phi_{1;3}^{(k)}$};
\draw[ultra thick,black,->] (13,16/3)--(13,22/3) node[midway,anchor=east] {$A^{\dagger}$};
\draw[black, ultra thick,->] (13,22/3)--(13,8);
\draw[red,dashed,very thick,->] (13,16/3)--(13,-1) node[midway,anchor=west] {$A$};
\end{tikzpicture}
\label{fig:Fokamoto}
\caption{\footnotesize{(Color online) Eigenfunctions $\phi_{n}^{(k)}$ of $H$ arranged according to the increasing eigenvalues $E_{n}$, together with the subspace decomposition $\mathcal{H}\equiv\mathcal{V}_{1}\oplus\mathcal{V}_{2}\oplus\mathcal{V}_{3}$ introduced in~\eqref{subspace}. The eigenfunctions $\phi_{n;j}^{(k)}$, for $j=1,2,3,$ and $n=0,1,\cdots$, are given in~\eqref{phin}, and computed through the three-term recurrence relation~\eqref{TTRR}.}}
\label{fig:spectrum}
\end{figure}

In this section, we construct the higher-modes through an alternative mechanism that relies on a three-term recurrence relation, similar to the usual construction for classical orthogonal polynomials~\cite{Chi78,Sze59}. To this end, we notice that, among all the zero-modes $\phi_{0,j}$, there is a common factor $e^{-\frac{x^{2}}{6}}/Q_{k+1}(x)$ that does not depend on the index $j$. We thus regard such a factor as the weight function $\mu_{k}(x)$ in the set of eigenfunctions of $H$ so that
\begin{equation}
\phi_{n;j}^{(k)}(x)=\frac{\mu_{k}(x)}{\mathcal{N}_{n,j}}P_{n;j}^{(k)}(x) \, , \quad \mu_{k}(x):=\frac{e^{-\frac{x^{2}}{6}}}{Q_{k+1}(x)} \, , \quad j=1,2,3,
\label{phin}
\end{equation}
with $\mathcal{N}_{n;j}$ the normalization constant, and $P_{n;j}^{(k)}(x)$ some functions to be determined from the following constraints:
\begin{equation}
P_{0;1}^{(k)}(x):=Q_{k}(x) \, , \quad P_{0;2}^{(k)}(x):=Q_{k+1,1}(x) \, , \quad P_{0;3}^{(k)}(x):=Q_{k+2,-1}(x) \, .
\label{initial}
\end{equation}
In this form, for $n=0$, we recover the zero-modes $\phi_{0;j}$ obtained in Sec.~\ref{subsec:zeromode}.The weight function $\mu_{k}(x)$ has support over the real line, $x\in\mathbb{R}$. It is worth to notice that the space of solutions $\mathcal{H}$ can be decomposed as the direct sum of the subspaces
\begin{equation}
\mathcal{H}=\mathcal{V}_{1}\oplus\mathcal{V}_{2}\oplus\mathcal{V}_{3} \, , \quad \mathcal{V}_{j}=\operatorname{Span}\{\phi_{n;j}\}_{n=0}^{\infty} \, , \quad j=1,2,3 \, ,
\label{subspace}
\end{equation}
where the ladder operators $A$ and $A^{\dagger}$ act as endomorphisms on the elements of the subspaces $\mathcal{V}_{j}$, that is, $A:\mathcal{V}_{j}\rightarrow\mathcal{V}_{j}$ and $A^{\dagger}:\mathcal{V}_{j}\rightarrow\mathcal{V}_{j}$ for all $j=1,2,3$, see Fig.~\ref{fig:spectrum}. 

We can exploit the latter fact to obtain additional properties by working in each vector subspace. In particular, from the eigenvalue equation~\eqref{H} we get a linear second-order differential equation for each set of functions $\{P^{(k)}_{n;j}\}_{n=0}^{\infty}$. It is useful to notice that the weight function $\mu_{k}$ satisfy
\begin{equation}
\frac{\mu_{k}'}{\mu_{k}}=-\frac{x}{3}-\frac{d}{dx}\ln Q_{k+1} \, , \quad \frac{\mu_{k}''}{\mu_{k}}=-\frac{1}{3}-\frac{d^{2}}{dx^{2}}\ln Q_{k+1}+\left(\frac{x}{3}+\frac{d}{dx}\ln Q_{k+1}\right)^{2} \, ,
\end{equation}
from which we determine the differential equation for the functions $P_{n;j}^{(k)}$ as
\begin{equation}
\frac{d^{2}P_{n;j}^{(k)}}{dx^{2}}-\left(\frac{2x}{3}+2\frac{Q_{k+1}'}{Q_{k+1}}\right)\frac{dP_{n;j}^{(k)}}{dx}+\left(\frac{Q_{k+1}''}{Q_{k+1}}+\frac{2x}{3}\frac{Q_{k+1}'}{Q_{k+1}}-\frac{4k}{3}+E_{n;j}\right)P_{n;j}^{(k)}=0 \, ,
\label{diffPn}
\end{equation}
for $ j=1,2,3$, and $n=0,1,\cdots$. 

%\textcolor{red}{I'M STILL EXPLORING WHETHER THE ALGEBRAIC-BETHE-ANSATZ COULD PROVIDE US WITH A PROOF ABOUT THE POLYNOMIAL STRUCTURE OF $P_{n;j}^{(k)}$. NEVERTHELESS, THE LATTER IS NOT SO NECESSARY FOR THE TIME BEING.}

%-------------->
%-------------->

%----------------------------------------------------

%----------------------------------------------------
\subsection{Three-term recurrence relation}
\label{subsec:TTRR}
Now, we present another of the main results of this work, that is, the proper construction of a linear \textit{three-term recurrence relation} for the polynomials $P_{n;j}^{(k)}$, for $j=1,2,3$, given in Eq.~\eqref{phin}. Such recurrence relations are fundamental in the theory of classical orthogonal polynomials~\cite{Chi78,Sze59}, a property contained within the \textit{Favard's theorem}~\cite{Fav35}, and in this case are achieved by exploiting the action of the third-order ladder operators $A$ and $A^{\dagger}$ on the elements of $\mathcal{V}_{j}$, for $j=1,2,3$, instead of the whole space $\mathcal{H}$. Thus, let us consider the action of the ladder operators $A$ and $A^{\dagger}$ on any arbitrary element of $\mathcal{V}_{j}$. Given that the ladder operators are mutually adjoint, we obtain 
\begin{equation}
A^{\dagger}\phi_{n;j}^{(k)}=C_{n+1;j}\phi_{n+1;j}^{(k)} \, , \quad A\phi_{n;j}^{(k)}=C_{n;j}\phi_{n-1;j}^{(k)} \, ,  
\label{Adagger}
\end{equation}
where $C_{n;j}$ is a proportionality constant computed from
\begin{equation}
(\phi_{n;j}^{(k)},AA^{\dagger}\phi_{n;j}^{(k)})=\vert C_{n+1;j}\vert^{2}.
\end{equation}
To determine the latter, we use the factorization $AA^{\dagger}=M_{1}^{\dagger}M_{2}^{\dagger}QQ^{\dagger}M_{2}M_{1}$ along with the intertwining relations~\eqref{intertH1M2}-\eqref{intertH2M1} and~\eqref{facI1I2}. We thus get $AA^{\dagger}=(H-\epsilon_{1})(H-\epsilon_{2})(H+2)$, and, with the aid of~\eqref{e1e2}, we obtain the proportionality constants as
\begin{align}
& C_{n+1;1}:=\sqrt{2^{3}(n+1)\left(n-k+\frac{2}{3}\right)\left(n-k+\frac{1}{3}\right)} \, , \\
& C_{n+1;2}:=\sqrt{2^{3}(n+1)\left(n+\frac{2}{3}\right)\left(n+k+\frac{4}{3}\right)} \, , \\
& C_{n+1;3}:=\sqrt{2^{3}(n+1)\left(n+\frac{4}{3}\right)\left(n+k+\frac{5}{3}\right)} \, .
\end{align}
From the latter, together with Eq.~\eqref{Adagger}, the normalization constants $\mathcal{N}_{n;j}$ are determined in terms of the normalization constants $\mathcal{N}_{0;j}$ of the corresponding zero-mode as
\begin{equation}
\mathcal{N}_{n+1;j}:=\mathcal{N}_{0;j}\prod_{p=0}^{n}C_{p+1;j} \, , \quad n=0,1,\cdots \, , \quad j=1,2,3.
\label{N123}
\end{equation}

Moreover, from the explicit action of the ladder operators $A$ and $A^{\dagger}$, combined with the differential equations obtained in Eq.~\eqref{diffPn}, we determine the following set of equations:
\begin{align}
& \mathcal{G}_{n;j}(x)\frac{dP_{n;j}^{(k)}}{dx}+\left[-\frac{Q_{k}'}{Q_{k}}\mathcal{G}_{n;j}(x)+E_{n;j}w_{k,0}^{[1]}\right]P_{n;j}^{(k)}=\mathcal{L}_{n;j}P_{n-1;j}^{(k)}  \, ,
\label{D1}\\
& \mathcal{G}_{n+1;j}(x)\frac{dP_{n;j}^{(k)}}{dx}-\left[\left(\frac{2x}{3}+\frac{Q_{k,1}'}{Q_{k,1}}\right)\mathcal{G}_{n+1;j}(x)+\left(\frac{2}{3}-2k+E_{n;j}\right)w_{k,0}^{[3]}\right]P_{n;j}^{(k)}=-\tilde{\mathcal{L}}_{n;j}P_{n+1;j}^{(k)}  \, ,
\label{D2}
\end{align}
where
\begin{align}
&\mathcal{G}_{n;j}(x):=\frac{2}{9}\frac{Q_{k+2}Q_{k}}{Q_{k+1}^{2}}-E_{n;j} \equiv\frac{2}{9}\frac{Q_{k+1,1}Q_{k+1,-1}}{Q_{k+1}^{2}}+\frac{2}{3}+2k-E_{n;j} \, , \\
&\mathcal{L}_{0;j}:=0 \, , \quad \mathcal{L}_{n+1;j}:=1 \, , \quad \tilde{\mathcal{L}}_{n;j}:=(C_{n+1;j})^{2} \, , \quad n=0,1,\cdots \, , \quad j=1,2,3\, .
\end{align}
By subtracting Eq.~\eqref{D1} with Eq.~\eqref{D2}, and after performing some calculations, we get
\begin{equation}
\tilde{\mathcal{L}}_{0;j}P_{1;j}^{(k)}(x)=\left[-w_{k,0}^{[2]}(x)\mathcal{G}_{1;j}(x)+E_{0;j}w_{k,0}^{[1]}(x)\frac{\mathcal{G}_{1;j}(x)}{\mathcal{G}_{0;j}(x)}+w_{k,0}^{[3]}(x)\left(\frac{2}{3}-2k+E_{0;j}\right)\right]P_{0;j}^{(k)}(x) \, ,
\label{P1}
\end{equation}
together with the three-term recurrence relation
\begin{multline}
\tilde{\mathcal{L}}_{n+1;j}P_{n+2;j}^{(k)}(x)+\frac{\mathcal{G}_{n+2;j}(x)}{\mathcal{G}_{n+1;j}(x)}P_{n;j}^{(k)}(x)\\
=\left[-w_{k,0}^{[2]}(x)\mathcal{G}_{n+2;j}(x)+E_{n+1;j}w_{k,0}^{[1]}(x)\frac{\mathcal{G}_{n+2;j}(x)}{\mathcal{G}_{n+1;j}(x)}+w_{k,0}^{[3]}(x)\left(\frac{2}{3}-2k+E_{n+1;j}\right)\right]P_{n+1;j}^{(k)}(x) \, ,
\label{TTRR}
\end{multline}
for $j=1,2,3$ and $n=0,1,\cdots$, where $w_{k,0}^{[p]}$ are given in Eqs.~\eqref{w1}-\eqref{w3} for $p=1,2,3$, and the initial conditions $P_{0;j}^{(k)}$ are given in~\eqref{initial}. In terms of finite-difference calculus, we notice that Eq.~\eqref{TTRR} is a second-order finite-difference equation that requires two boundary conditions to determine $P_{n;j}^{(k)}$ uniquely. However, for $n=0$, we have $L_{0;j}=0$ and thus $P^{(k)}_{1;j}$ is determined exclusively from $P^{(k)}_{0;j}$, for $j=1,2,3$. The latter means that we only require $P^{(k)}_{0,j}$ as the initial condition for each polynomial sequence, the remaining terms are then computed recursively. To illustrate our results, we present in Tables~\ref{pnk}-\ref{pnk2} the explicit form of $P_{n,j}^{(k)}$ for several values of $k$ and $n$. Such terms are computed using the generalized Okamoto polynomials given in Tables~\ref{tabQk}-\ref{tabQkm1}. 

From the three-term recurrence relation in~\eqref{TTRR}, it seems that $P_{n;j}^{(k)}$ takes the form of a rational function. Nevertheless, after computing explicitly some terms, it follows that $P_{n,j}^{(k)}$ leads indeed to polynomials. Such a computation is presented in Tables~\ref{pnk}-\ref{pnk2}, for several values of $n$ and $k=1,3$. Clearly, the latter is by no means a proof about the polynomial structure of $P_{n;k}^{(k)}$, and a more detailed analysis is required. This is a fact discussed in the following sections.

%-------->
\begin{table}
\centering
\footnotesize{\begin{tabular}{|c|p{7cm}|p{7cm}|}
\hline
\rowcolor{gray} $\textcolor{white}{n}$ & $\textcolor{white}{P_{n;1}^{(k=1)}(x)/\mathcal{D}_{n;1}^{(k=1)}}$ & $\textcolor{white}{P_{n;1}^{(k=3)}/\mathcal{D}_{n;1}^{(k=3)}}$ \\
\hline
 0 & $1$ & $8 x^6+60 x^4+90 x^2+135$ \\
 \hline
 1 & $x(2 x^2+9)$ & $x(16 x^8+480 x^6+3528 x^4+7560 x^2+8505)$ \\
 \hline
 2 & $8 x^6-36 x^4-162 x^2+81$ & $64 x^{12}+2496 x^{10}+35280 x^8+211680 x^6+510300 x^4+510300 x^2-382725$\\
 \hline
  3 & $x(16 x^8-432 x^6+1944 x^4+4860 x^2-10935)$ & $x(128 x^{14}+4416 x^{12}+58464 x^{10}+390960 x^8+1370520 x^6+2143260 x^4-2296350 x^2-10333575 )$ \\
  \hline
  4 & $64 x^{12}-4032 x^{10}+73872 x^8-381024 x^6-306180 x^4+2755620 x^2-688905$ & $512 x^{18}+8448 x^{16}-32256 x^{14}-1126656 x^{12}-6516288 x^{10}-12247200 x^8-7348320 x^6+165337200 x^4+310007250 x^2-93002175$\\
  \hline
%  5 & $x(128 x^{14}-14400 x^{12}+557280 x^{10}-8910000 x^8+52925400 x^6-30311820 x^4-378897750 x^2+341007975)$ & $x(1024 x^{20}-15360 x^{18}-472320 x^{16}+1520640 x^{14}+48988800 x^{12}+165628800 x^{10}-376164000 x^8-1417176000 x^6-11780275500 x^4-9300217500 x^2+20925489375)$\\
%\hline
\rowcolor{gray} $\textcolor{white}{n}$ & $\textcolor{white}{P_{n;2}^{(k=1)}(x)/\mathcal{D}_{n;2}^{(k=1)}}$ & $\textcolor{white}{P_{n;2}^{(k=3)}/\mathcal{D}_{n;2}^{(k=3)}}$ \\
\hline
 0 & $4 x^4+12 x^2-9$ & $256 x^{16}+7680 x^{14}+80640 x^{12}+362880 x^{10}+453600 x^8-1905120 x^6-14288400 x^4-21432600 x^2+8037225$ \\
\hline
 1 & $x \left(8 x^6-84 x^4-126 x^2+567\right)$ & $x(512 x^{18}+3840 x^{16}-129024 x^{14}-1451520 x^{12}-2358720 x^{10}+24766560 x^8+106142400 x^6+445798080 x^4+208967850 x^2-940355325 )$ \\
\hline
 2 & $32 x^{10}-1200 x^8+10080 x^6-85050 x^2+25515$ & $2048 x^{22}-58368 x^{20}-806400 x^{18}+13996800 x^{16}+131155200 x^{14}-240589440 x^{12}-4606580160 x^{10}-8438320800 x^8-39285955800 x^6+56421319500 x^4+197474618250 x^2-42315989625$\\
\hline
  3 & $x(64 x^{12}-4992 x^{10}+121680 x^8-1010880 x^6+1326780 x^4+7960680 x^2-8955765)$ & $x(4096 x^{24}-319488 x^{22}+4773888 x^{20}+85432320 x^{18}-1369094400 x^{16}-8770083840 x^{14}+58205226240 x^{12}+430768316160 x^{10}-186041091600 x^8+1715208112800 x^6-14150466930600 x^4-19296091269000 x^2+21708102677625)$\\
\hline
  4 & $256 x^{16}-33792 x^{14}+1587456 x^{12}-32617728 x^{10}+283551840 x^8-700539840 x^6-1576214640 x^4+4728643920 x^2-886620735$ & $16384 x^{28}-2310144 x^{26}+102371328 x^{24}-1135337472 x^{22}-22142647296 x^{20}+383543424000 x^{18}+1297481898240 x^{16}-21340876170240 x^{14}-79981334072640 x^{12}+351830022223680 x^{10}-222364480338000 x^8+5263973698183200 x^6+955156517815500 x^4-14327347767232500 x^2+2149102165084875$\\
%  5 & $512 x^{19}-102144 x^{17}+7792128 x^{15}-289578240 x^{13}+5512328640 x^{11}-51022651680 x^9+179688468960 x^7+89844234480 x^5-1179205577550 x^3+758060728425 x$ & $-382725 + 510300 x^2 + 510300 x^4 + 211680 x^6 + 35280 x^8 +   2496 x^{10} + 64 x^{12}$\\
\hline
\end{tabular}}
\caption{Polynomials $P_{n;j}^{(k)}$ computed through the three-term recurrence relation~\eqref{TTRR} for $j=1,2$ and $n=0,1,2,3,4$. The constants $\mathcal{D}_{n;1}^{(k)}$ and $\mathcal{D}_{n;2}^{(k)}$ are integer scaling factors.}
\label{pnk}
\end{table}
%-------->

%-------->
\begin{table}[t]
\centering
\footnotesize{\begin{tabular}{|c|p{7cm}|p{7cm}|}
\hline
\rowcolor{gray} $\textcolor{white}{n}$ & $\textcolor{white}{P_{n;3}^{(k=1)}(x)/\mathcal{D}_{n;3}^{(k=1)}}$ & $\textcolor{white}{P_{n;3}^{(k=3)}/\mathcal{D}_{n;3}^{(k=3)}}$ \\
\hline
 0 & $x(4 x^4-45)$ & $x(256 x^{16}+6144 x^{14}+34560 x^{12}-138240 x^{10}-2138400 x^8-8553600 x^6-22453200 x^4+63149625)$ \\
\hline
 1 & $16 x^8-288 x^6+360 x^4+3240 x^2-1215$ & $1024 x^{20}-3072 x^{18}-407808 x^{16}-1797120 x^{14}+18506880 x^{12}+155675520 x^{10}+314344800 x^8+808315200 x^6-2778583500 x^4-6820159500 x^2+1705039875$ \\
 \hline
 2 & $x(32 x^{10}-1584 x^8+20592 x^6-49896 x^4-187110 x^2+280665 )$ & $x(2048 x^{22}-89088 x^{20}-290304 x^{18}+27530496 x^{16}+66624768 x^{14}-1684126080 x^{12}-8355856320 x^{10}+6870679200 x^8+1717669800 x^6+378746190900 x^4+405799490250 x^2-608699235375)$\\
\hline
 3 & $128 x^{14}-12096 x^{12}+376992 x^{10}-4490640 x^8+15717240 x^6+23575860 x^4-106091370 x^2+22733865$ & $8192 x^{26}-798720 x^{24}+19353600 x^{22}+102574080 x^{20}-5769792000 x^{18}-629233920 x^{16}+391975718400 x^{14}+884443795200 x^{12}-6959998029600 x^{10}-3246395922000 x^8-95444040106800 x^6+14608781649000 x^4+295827828392250 x^2-49304638065375$\\
\hline
 4 & $x(256 x^{16}-39168 x^{14}+2193408 x^{12}-56137536 x^{10}+661620960 x^8-2977294320 x^6+20096736660 x^2-15072552495)$ & $x(16384 x^{28}-2703360 x^{26}+149409792 x^{24}-2765905920 x^{22}-15309388800 x^{20}+819467228160 x^{18}-1871143545600 x^{16}-54521712645120 x^{14}+22612029854400 x^{12}+1482481454126400 x^{10}-1026139373859600 x^8+12480073465860000 x^6-27518561992221300 x^4-58968347126188500 x^2+44226260344641375)$\\
% 5 & $\frac{8 x^6-36 x^4-162 x^2+81}{1620}$ & $-382725 + 510300 x^2 + 510300 x^4 + 211680 x^6 + 35280 x^8 + 2496 x^{10} + 64 x^{12}$\\
\hline
\end{tabular}}
\caption{Polynomials $P_{n;j}^{(k)}$ computed through the three-term recurrence relation~\eqref{TTRR} for $j=3$ and $n=0,1,2,3,4$.  The constants $\mathcal{D}_{n;3}^{(k)}$ are integer scaling factors.}
\label{pnk2}
\end{table}
%-------->

%----------------------------------------------------

%------------------------------------------------------->
\section{Connection to higher-order SUSY QM}
\label{sec:SUSY}
So far, we have identified the potential $V^{(k)}(x)$ written in terms of the Okamoto polynomials and constrained to the third-order shape-invariant condition~\eqref{shapeH}. In contradistinction to previous works, the spectral problem was determined in general for any value of the arbitrary parameter $k=0,1,\cdots$. Moreover, a mechanism to generate the higher-modes through a three-term recurrence relation, for arbitrary $k$, was introduced. In this section, for completeness, we discuss the connection from our previous results to the higher-order SUSY QM construction. It is worth to mention that, for the construction discussed in~\cite{Mar09}, it was found that the resulting potential was related to a two-step SUSY construction of the harmonic oscillator. Such a case corresponds to $k=1$ in our construction. We thus complement such a discussion, and generalize it for any values of $k$.

Before proceeding, we explore the asymptotic behavior of the rational term $\frac{Q_{k+2}Q_{k}}{Q_{k+1}^{2}}$ in the potential $V^{(k)}(x)$ obtained in~\eqref{potVk}, where $Q_{k}\equiv Q_{k,0}$,. It is well-known~\cite{Oka86,Cla08} that $Q_{m,n}$ is a polynomial of degree $d_{m,n}=m^{2}+n^{2}+mn-m-n$, and thus the rational part of the potential $V^{(k)}$ is a polynomial of degree $2(k^{2}+k+1)$ and $2k(k+1)$ in the numerator and denominator, respectively. The latter implies that the whole potential $V^{(k)}(x)$ behaves asymptotically as a polynomial of degree two for $x\rightarrow\pm\infty$. Thus, from Oblomkov's theorem~\cite{Obl99}, it follows that $V^{(k)}(x)$ can be constructed from the harmonic oscillator through several rational Darboux transformations.

The generalized Okamoto polynomials admit a Wronskian representation obtained from the relationship between the Schur polynomials and the $\tau$ function defining the Toda-type equations~\eqref{recurrenceQ1}-\eqref{recurrenceQ2}. See~\cite{Nou99,Cla06,Kaj98} for further details. Such a Wronskian representation allows us identifying the corresponding elements required to perform the Darboux transformation to generate $V^{(k)}(x)$. Thus, from~\cite{Cla08} we have
\begin{equation}
Q_{m,n}=C_{m,n}\operatorname{Wr}(\Psi_{1},\cdots,\Psi_{3m+3n-5},\Psi_{2}, \cdots,\Psi_{3n-4}) \, , \quad n=1,2,\cdots \, , \quad m=0,1,\cdots \, ,
\label{wronskian1}
\end{equation}
where the proportionality constant $C_{m,n}$ is fixed so that~\eqref{recurrenceQ1}-\eqref{recurrenceQ2} are fulfilled, and $\operatorname{Wr}(\cdot)$ denotes the usual Wronskian determinant
\begin{equation}
\operatorname{Wr}(f_{1},f_{2},\cdots,f_{n})=\left\vert\begin{array}{cccc}
f_{1} & f_{2} & \cdots & f_{n} \\
f'_{1} & f'_{2} & \cdots & f'_{n} \\
\vdots & \vdots & \ddots & \vdots \\
f^{(n-1)}_{1} & f^{(n-1)}_{2} & \cdots & f^{(n-1)}_{n}
\end{array} \right\vert \, , \quad f^{(n)}\equiv\frac{d^{n}f}{dx^{n}} \, .
\label{Wr}
\end{equation}
The functions $\psi_{n}(x)$ are determined from the generating function~\cite{Cla08}
\begin{equation}
e^{2x\xi+3\xi^{2}}=\sum_{r=0}^{\infty}\Psi_{r}(x)\xi^{r} \, , \quad \Psi_{r}(x):=\frac{1}{3^{r/2}r!}\mathcal{H}_{r}\left(\frac{x}{\sqrt{3}}\right) \, ,
\label{psir}
\end{equation}
with $\mathcal{H}_{r}(z)=i^{r}\mathtt{H}_{r}(iz)$ and $\mathtt{H}_{r}(z)$  the \textit{pseudo-Hermite} and \textit{Hermite} polynomials~\cite{Olv10}, respectively. Notice that $C_{m,n}$ does not modify the form of $V^{(k)}(x)$ as such a constant vanishes inside the logarithmic derivative. Moreover, $C_{m,n}$ can be absorbed by the zero-modes normalization factor, and thus its explicit form is not required throughout the rest of the manuscript.

The Wronskian representation for the conventional Okamoto polynomials $Q_{m}\equiv Q_{m,0}$ must be handled with caution, as the length of the partition defining the corresponding $\tau$ function used in~\eqref{wronskian1} is not the appropriate one, see~\cite{Nou99}. In such a case we get
\begin{equation}
Q_{m}=C_{m}\operatorname{Wr}(\Psi_{2}, \cdots,\Psi_{3m-4}) \, , \quad m=0,1,\cdots \, .
\label{wronskian0}
\end{equation}
Notice that fixing $n=0$ in~\eqref{wronskian1} does not lead to the right representation given in~\eqref{wronskian0}.

On the other hand, the generalized Okamoto polynomials $Q_{m,n}$ possess a symmetry~\cite{Cla08} of the form $Q_{m,n}(x)=\exp(-\frac{i}{2}\pi d_{m,n})Q_{n,m}(ix)$, with $d_{m,n}=m^{2}+n^{2}+mn-m-n$ the degree of the Okamoto polynomial. This symmetry allows us getting the alternative representation 
\begin{equation}
Q_{m,n}=D_{m,n}\operatorname{Wr}(\psi_{1},\psi_{4},\cdots \psi_{3n+3m-5},\psi_{2},\psi_{5},\cdots \Psi_{3m-4})  \, , 
\label{wronskian2}
\end{equation}
where
\begin{equation}
\psi_{r}(x):=(-i)^{r}\Psi_{r}(-ix)=\frac{1}{3^{r/2}r!}\texttt{H}_{r}\left(\frac{x}{\sqrt{3}}\right) \, ,
\label{psir1}
\end{equation}
and $D_{m,n}$ is a proportionality constant. In this form, we have a Wronskian representation in terms of pseudo-Hermite polynomials~\eqref{wronskian1} and one in terms of Hermite polynomials~\eqref{wronskian2}. Additional representations have been recently reported in terms of \textit{pseudo-Wronskians}, which are composed of a combination of Hermite and pseudo-Hermite polynomials~\cite{Gom18}. During the rest of this manuscript, we focus mostly on the representation~\eqref{wronskian2} as it reproduces the right results for $n=0$ and $n=-1$.

The Wronskian representations allow finding a relation between the potential $V^{(k)}(x)$ and the higher-order Darboux transformations. To this end, let us introduce the following short and convenient notation:
\begin{equation}
\mathcal{W}_{\{i_{1},\cdots,i_{m}\}}=\operatorname{Wr}\left(\psi_{i_{1}},\cdots,\psi_{i_{m}} \right) \, , \quad \mathfrak{W}_{\{i_{1},\cdots,i_{m}\}}=\operatorname{Wr}\left(\Psi_{i_{1}},\cdots,\Psi_{i_{m}} \right) \, ,
\label{wronskiantilde0} 
\end{equation}
together with 
\begin{align}
& \tilde{\mathcal{W}}_{\{i_{1},\cdots,i_{m}\}}=\operatorname{Wr}\left(\tilde{\psi}_{i_{1}},\cdots,\tilde{\psi}_{i_{m}} \right) \, , && \tilde{\mathfrak{W}}_{\{i_{1},\cdots,i_{m}\}}=\operatorname{Wr}\left(\tilde{\Psi}_{i_{1}},\cdots,\tilde{\Psi}_{i_{m}} \right) \, ,
\label{wronskiantilde1} \\
& \tilde{\psi}_{r}:=e^{-\frac{x^{2}}{6}}\psi_{r}=\frac{e^{-\frac{x^{2}}{6}}}{3^{r/2}r!}\mathtt{H}_{r}\left(\frac{x}{\sqrt{3}}\right) \, , && \tilde{\Psi}_{r}:=e^{\frac{x^{2}}{6}}\Psi_{r}=\frac{e^{\frac{x^{2}}{6}}}{3^{r/2}r!}\mathcal{H}_{r}\left(\frac{x}{\sqrt{3}}\right) \, .
\label{wronskiantilde2}
\end{align}
With this new notation, the conventional Okamoto polynomials $Q_{k+1}$ can be rewritten as
\begin{align}
&Q_{k+1}=C_{k+1}\mathcal{W}_{\{1,\cdots, 3k-2,2,\cdots,3k-1\}}=C_{k+1}\, e^{\frac{k}{3}x^{2}} \, \tilde{\mathcal{W}}_{\{1,\cdots,3k-2,2,\cdots,3k-1\}} \, , \\
&Q_{k+1}=D_{k+1}\mathfrak{W}_{\{2,\cdots,3k-1\}}=D_{k+1} \, e^{-\frac{k}{6}x^{2}} \, \tilde{\mathfrak{W}}_{\{2,\cdots,3k-1\}} \, ,
\end{align}
where we have used the Wronskian identity Wr$(g f_{1},\cdots, g f_{n})=g^{n}$Wr$(f_1,\cdots,f_n)$~\cite{Oda20}.

Therefore, the potential $V^{(k)}(x)$ given in~\eqref{potVk2} takes the two equivalent forms
\begin{align}
V^{(k)}(x)&\equiv\frac{x^{2}}{9}-2\frac{d^{2}}{dx^{2}}\ln \tilde{\mathcal{W}}_{\{1,4,\cdots,3k-2;2,5,\cdots3k-1\}}-\frac{1}{3}  \, ,
\label{potWr1}\\
V^{(k)}(x)& \equiv\frac{x^{2}}{9}-2\frac{d^{2}}{dx^{2}}\ln \tilde{\mathfrak{W}}_{\{2,5,\cdots3k-1\}}+2k-\frac{1}{3} \, .
\label{potWr2}
\end{align}
These expressions reveal that the third-order shape invariant construction~\eqref{shapeH} is related to a higher-order SUSY transformation. In particular, \eqref{potWr1} shows that $H$ given in~\eqref{shapeH} is related to a state-deleting SUSY transformation of the harmonic oscillator (See App.~\ref{sec:APPC} for details) with the physical energy levels $\{E_{n}^{(osc)}\}_{n\in\mathcal{S}_{d}}$ deleted, where $\mathcal{S}_{d}=\{ 1,2,4,5,\cdots 3k-2,2k-1 \}$. Alternatively, from the Wronskian representation~\eqref{potWr2}, we obtain a state adding construction in which the non-physical energy levels $\{E_{-n}^{(osc)}\}_{n\in\mathcal{S}_{a}}$ have been added, with $\mathcal{S}_{a}=\{ 2,5,\cdots 3k-1\}$. In this form, we verify the well-known equivalence between the state-adding and the state-deleting Darboux transformations already noticed by Felder \textit{et al.}~\cite{Fel12}.

Lastly, using~\eqref{wronskian1} and the action of the creation operator $A^{\dagger}$, the higher-modes can be alternatively determined from the Wronskian representation, leading to (see App.~\ref{sec:APPB} for details)
\begin{equation}
\begin{aligned}
&\phi_{n;1}^{(k)}(x)\equiv\frac{\mu_{k}(x)}{\sqrt{\tilde{\mathcal{N}}_{n;1}}} \mathcal{W}_{I^{(k)}\cup\{3n\}}  \, , \quad 
\phi_{n;2}^{(k)}(x)\equiv\frac{\mu_{k}(x)}{\sqrt{\tilde{\mathcal{N}}_{n;2}}} \mathcal{W}_{I^{(k)}\cup\{3n+3k+1\}} \, , \\ 
&\phi_{n;3}^{(k)}(x)\equiv\frac{\mu_{k}(x)}{\sqrt{\tilde{\mathcal{N}}_{n;3}}} \mathcal{W}_{I^{(k)}\cup\{3n+3k+2\}} \, , 
\end{aligned}
\label{highermodesW}
\end{equation}
with $I^{(k)}=\{1,2,4,5,\cdots,3k-2,3k-1\}$, $\mu_{k}(x)$ the weight function given in~\eqref{phin}, and $\tilde{\mathcal{N}}_{n;j}$ the corresponding normalization constants, which are different to the ones obtained in~\eqref{N123}. We thus have two different forms to represent the higher-modes. Furthermore, after comparing~\eqref{phin} with~\eqref{highermodesW}, it is straightforward to realize that 
\begin{equation}
\begin{aligned}
& P_{n;1}^{(k)}(x)=\sqrt{\frac{\tilde{\mathcal{N}}_{n;1}}{\mathcal{N}_{n;1}}} \, \mathcal{W}_{I^{(k)}\cup\{3n\}}(x) \, , \quad P_{n;2}^{(k)}(x) = \sqrt{\frac{\tilde{\mathcal{N}}_{n;2}}{\mathcal{N}_{n;2}}} \mathcal{W}_{I^{(k)}\cup\{3n+3k+1\}}(x) \, , \\ 
& P_{n;3}^{(k)}(x) = \sqrt{\frac{\tilde{\mathcal{N}}_{n;3}}{\mathcal{N}_{n;3}}} \mathcal{W}_{I^{(k)}\cup\{3n+3k+2\}}(x) \, ,
\end{aligned}
\label{PW}
\end{equation}
for $n,k=0,1,\cdots$. From the previous Wronskian representation, it is clear that the functions $P_{n;j}^{(k)}$ computed from~\eqref{TTRR} are indeed polynomials.

\subsection{Relation with the exceptional Hermite polynomials}
The Wronskian representation obtained in the previous section provides us with an additional relation that allows relating the polynomials emerging from the higher-modes with the family of orthogonal \textit{exceptional Hermite polynomials}~\cite{Gom14}. To this end, let us introduce the non-decreasing sequence of integers, $0\leq\lambda_{1}\leq\cdots\leq\lambda_{\ell}$, together with the partition and double partition
\begin{equation}
\lambda:=(\lambda_{1},\cdots, \lambda_{\ell}) \, , \quad \lambda^{2}:=(\lambda_{1},\lambda_{1},\cdots, \lambda_{\ell}, \lambda_{\ell}) \, ,
\label{partition}
\end{equation}
respectively. The exceptional Hermite polynomials are defined as~\cite{Gom14}
\begin{equation}
H_{n}^{(\lambda)}(x)\equiv H_{\lambda^{2},n}(x):=\operatorname{Wr}[\overline{\texttt{H}}_{\nu_1},\overline{\texttt{H}}_{\nu_1+1},\cdots \overline{\texttt{H}}_{\nu_{\ell}},\overline{\texttt{H}}_{\nu_{\ell}+1},\overline{\texttt{H}}_{n}] \, , \quad n\not\in\{\nu_{1},\nu_{1}+1,\cdots, \nu_{\ell},\nu_{\ell}+1\} \, ,
\label{Hlj}
\end{equation}
where $\overline{\texttt{H}}_{m}:=\texttt{H}_{m}(x)$ and the indexes inside the Wronskian are related to the elements of the partition through
\begin{equation}
\nu_{j}=\lambda_{j}+j-1 \, , \quad 0\leq \nu_{1}< \cdots< \nu_{\ell} \, , \quad j=1,\cdots \ell.
\label{lambdak}
\end{equation}
From the latter, it is clear that the polynomials $P_{n;j}^{(k)}$ becomes a particular case of the exceptional Hermite polynomials~\eqref{Hlj}. It is worth noticing that the argument of the Hermite polynomials inside the Wronskian~\eqref{PW} is different to that of the Hermite polynomials in the Wronskian~\eqref{Hlj}. Nevertheless, a simple reparametrization of the form $x\rightarrow \sqrt{3}x$ solves the issue, leading to the relation
\begin{equation}
H_{\sigma_{n;j}}^{(\lambda_{k})}(x)\equiv \mathfrak{D}_{n;j}^{(k)}\mathcal{W}_{\mathcal{I}_{1}\cup\{\sigma_{n;j}\}}(\sqrt{3}x)=\mathfrak{D}_{n;j}^{(k)}\sqrt{\frac{\mathcal{N}_{n;j}}{\tilde{\mathcal{N}}_{n;k}}}P_{n;j}^{(k)}(\sqrt{3}x)  \, , \quad n=0,1,\cdots \, ,
\label{Hlambdak}
\end{equation}
where we have used the index notation
\begin{equation}
\sigma_{n;1}:=3n \, , \quad \sigma_{n;2}:=3n+3k+1 \, , \quad \sigma_{n;3}=3n+3k+2 \, .
\end{equation}
with the double partition $\lambda^{2}_{k}:=(1,1,2,2,\cdots,k,k)$, or equivalently the partition $\lambda_{k}:=(1,2,\cdots,k)$, so that $\nu=\{\nu_{p}\}_{p=1}^{2k}=\{1,2,4,5,\cdots,3k-2,3k-1\}$, for $k=0,1,\cdots$. On the other hand, the proportionality constant $\mathfrak{D}_{n;j}^{(k)}$ becomes
\begin{equation}
\mathfrak{D}_{n;j}^{(k)}:=3^{\frac{k(k+1)}{4}+\frac{3k^{2}}{2}+\frac{\sigma_{n;j}}{2}}\left(\prod_{p=1}^{k}\nu_{p}!\right)(\sigma_{n;j}!) \, .
\end{equation}

Recursion formulas have been previously constructed for the families of exceptional Hermite polynomials~\cite{Oda13,Gom14,Gom16} through the use of higher-order recurrence relations. In particular, for any family of general exceptional Hermite polynomial $H_{n}^{(\lambda)}$ defined in terms of an $(\ell+1)$-order Wronskian~\eqref{Hlj}, there is a $(2\ell+3)$-order recurrence relation, with $(\ell+1)$ initial conditions, that generates the complete sequence of exceptional polynomials. See~\cite{Gom14} for more details. From the latter, it follows that the particular sequence $\lambda_{k}$ defining $H_{\sigma_{n;j}}^{(\lambda_k)}$ in~\eqref{Hlambdak} should be computed from a $(2k+3)$-order recurrence relation. Nevertheless, from the recursion formulas ~\eqref{TTRR} obtained for the polynomials $P_{n;j}^{(k)}$, we can introduce a set of new linear and third-order recurrence relations for the exceptional Hermite polynomials defined by the three disjoint sets $S_{H_{1}}:=\{H_{\sigma_{n;1}}^{(\lambda_{k})}\}_{n=0}^{\infty}$, $S_{H_{2}}:=\{H_{\sigma_{n;2}}^{(\lambda_{k})}\}_{n=0}^{\infty}$, and $S_{H_{3}}:=\{H_{\sigma_{n;3}}^{(\lambda_{k})}\}_{n=0}^{\infty}$. That is, the rescaled exceptional Hermite polynomials defined in~\eqref{Hlambdak} satisfy a three-term recurrence relation akin to that of orthogonal polynomials. 

Thus, we can reduce the problem of the $(2k+3)$-order recurrence relation to three third-order recurrence relations, and the number of initial conditions are reduced from $(k+1)$ to only one for each sequence.

%The normalization constants $\mathcal{N}^{(k)}_{n;1}$ are (see App.~\ref{sec:APPB} for details)
%\begin{equation}
%\begin{aligned}
%& \mathcal{N}_{n;1}=\sqrt{\pi}\,2^{2k}\left(\frac{1}{3}-n\right)_{k}\left(\frac{2}{3}-n\right)_{k} \, , \quad
%\mathcal{N}_{n;2}=\sqrt{\pi}\,2^{2k}\left(n+1\right)_{k}\left(n+\frac{2}{3}\right)_{k} \, , \\ 
%& \mathcal{N}_{n;3}=\sqrt{\pi}2^{2k}\left(n+1\right)_{k}\left(n+\frac{4}{3}\right)_{k} \, .	
%\end{aligned}
%\end{equation}

%----------------------------------------------------

\subsection{Number of zeros}
\label{subsec:Wzeros}
As previously mentioned, the eigenvalue problem associated with the Hamiltonian $H$ in~\eqref{H} is equivalent to a Sturm-Liouville problem. In consequence, the eigenfunctions should satisfy the well-known \textit{oscillation theorem}~\cite{Inc56}. In this section, we identify the number of zeros of the generalized Okamoto polynomials $Q_{m,n}$ and the polynomials $P_{n;j}^{(k)}$ in each sequence of solutions through the Wronskian representation. Such zeros are distributed over the support of the weight function $\mu_{k}(x)$, which in this case corresponds to the real line. Studies in this regard have been addressed by Felder~\cite{Fel12} for Wronksians defined in terms of sequences of contigous Hermite polynomials, and recently in~\cite{Gar15} for general sequences. In the latter, the authors conjectured that a Wronskian of the form Wr$(\mathtt{H}_{\xi_{1}},\cdots, \mathtt{H}_{\xi_{\ell}})$ characterized by the partition $\tilde{\lambda}=(\tilde{\lambda}_{1},\cdots,\tilde{\lambda}_{\ell})$, with $\tilde{\lambda}_{j}=\xi_{j}-j+1$, has the following number of real zeros~\cite{Gar15}:
\begin{itemize}
\item[(a)] A zero of multiplicity $n_{0}=\frac{d_{\xi}(d_{\xi}+1)}{2}$ at $x=0$. With $d_{\xi}=p-q$, where $p$ and $q$ are the number of odd and even elements, respectively, in the sequence $\{\xi_{1},\cdots,\xi_{\ell}\}$.
\item[(b)] $n_{+}$ simple positive zeros, where
\begin{equation}
n_{+}:=\frac{1}{2}\left( \sum_{j=1}^{\ell}(-1)^{\ell-j} \tilde{\lambda}_{j}-\frac{\left\vert d_{\xi}+(\ell-2\lfloor\frac{\ell}{2}\rfloor)\right\vert}{2} \right)
\label{conjecture}
\end{equation}
\item[(c)] The same number of simple negative zeros, $n_{-}=n_{+}$,
\end{itemize}
where $0\leq\xi_{1}<\cdots\xi_{ell}$ and $0\leq\tilde{\lambda}_{1}\leq\cdots\leq\tilde{\lambda}_{\ell}$.

In particular, considering the generalized Okamoto polynomials representation~\eqref{wronskian1}, with $m,n=0,1,\cdots$ and $m+n=2,3,\cdots$, we identify $\ell=n+2m-2$ and the total number of real zeros $n_{t}=n_{0}+2n_{+}$ as shown in Tab.~\ref{ntQmn}. The latter reveals that the conventional Okamoto polynomials $Q_{k}\equiv Q_{k,0}$ are nodeless functions, as stated in Sec.~\ref{sec:thirdSI}.

\begin{table}
\centering
\begin{tabular}{|c|c|c|c|}
\hline
\rowcolor{gray} $\textcolor{white}{Q_{m,n}}$ & $\textcolor{white}{n_{0}}$ & $\textcolor{white}{n_{+}}$ & $\textcolor{white}{n_{t}=n_{0}+2n_{+}}$ \\
\hline
 $Q_{2m,2n}$ & $0$ & $n$ & $2n$ \\
 \hline
 $Q_{2m,2n+1}$ & $0$ & $n+m$ & $2(n+m)$ \\
 \hline
 $Q_{2m+1,2n}$ & $0$ & $n$ & $2n$ \\
 \hline
 $Q_{2m+1,2n+1}$ & $1$ & $n$ & $2(n+m)+1$ \\
\hline
\end{tabular}
\caption{Number of zeros at $x=0$ ($n_{0}$), $x>0$ ($n_{+}$), and the total number of zeros in $x\in\mathbb{R}$ ($n_{t}$) of the generalized Okamoto polynomials $Q_{m,n}$.}
\label{ntQmn}
\end{table}

On the other hand, we can identify the zeros of the polynomials $P_{n;j}^{(k)}$, which also are the zeros of the eigenfunctions $\phi_{n;j}^{(k)}$. This is achieved with the aid of the Wronskian representation~\eqref{PW} and ~\eqref{conjecture}. After some calculations, we obtain the total number of zeros of the eigenfunctions shown in Tab.~\ref{ntphi}. From the latter, it is clear that $\phi_{0;1}^{(0)}$ is always a nodeless function for all $n$ and $k$, and thus it corresponds to the ground state of the system. Moreover, the zero-modes $\phi_{0;2}^{(k)}$ and $\phi_{0;3}^{(k)}$ correspond to the $(k+1)$ and $(k+2)$ excited states of the system, respectively. 

\begin{table}
\centering
\begin{tabular}{|c|c|c|c|}
\hline
\rowcolor{gray} $\textcolor{white}{n}$ & $\textcolor{white}{\phi_{n;1}^{(k)}}$ & $\textcolor{white}{\phi_{n;2}^{(k)}}$ & $\textcolor{white}{\phi_{n;3}^{(k)}}$ \\
\hline
$0\leq n \leq k$ & $n$  & \multirow{2}{*}{$3n+k+1$} & \multirow{2}{*}{$3n+k+2$} \\ \cline{1-2}
$n>k$ & $3n-2k$  &                   &                   \\ \hline
\end{tabular}
\caption{Total number of zeros in $x\in\mathbb{R}$ of the eigenfunctions $\phi_{n;1}^{(k)}$, $\phi_{n;2}^{(k)}$, and $\phi_{n;3}^{(k)}$.}
\label{ntphi}
\end{table}
%-------->

%----------------------------------------------------
\section{Conclusions}
The identification of third-order shape-invariant Hamiltonians associated with rational solutions of the fourth Painlev\'e transcendent in the ``$-2x/3$'' hierarchy has been developed in the most general case so that the resulting potentials are free of any singularities. This condition leads to a family of potentials written in terms of the conventional Okamoto polynomials and three zero-modes constructed as the product of generalized Okamoto polynomial times a weight function with support on the real line. In this form, the results of this manuscript extend and generalize some previous statements made in earlier works~\cite{Mar09,Zel20c}. Interestingly, the existence of third-order ladder operators leads naturally to three sequences of eigenfunctions whose eigenvalues do not overlap. Although the eigenvalues of the shape-invariant Hamiltonian are not equidistant, the three independent sequences define sets of eigenvalues that are indeed equidistant within each sequence. The latter is a fundamental property exploited throughout the manuscript to properly identify a second-order differential equation, along with a second-order finite-difference equation for the eigenfunctions within each sequence. The coefficients of the finite-difference equation are such that only one initial condition per sequence, fixed as the corresponding zero-mode, is required to determine the higher-modes through finite iterations. This is a property akin to that of classical orthogonal polynomials~\cite{Chi78,Sze59}. The calculations presented throughout the manuscript were achieved with the aid of the new set of identities introduced in~\eqref{w1r}-\eqref{w3r}, determined from the appropriate B\"acklund transformations.

On the other hand, a direct relation between our results and that of higher-order Darboux-Crum (higher-order SUSY) transformations~\cite{Cru55} of the harmonic oscillator is established. In this form, it is found that the potentials constructed from the third-order shape-invariant condition and the ``$-2x/3$'' hierarchy of rational solutions are equivalent to $2k$ state-deleting Darboux transformation. Such a equivalence leads simultaneously to  a relation between the higher-modes and the family of exceptional Hermite polynomials, defined by the appropriate partition $\lambda_{k}$. Remarkably, the latter provides us with a set of three different three-term recurrence relations for the exceptional Hermite polynomials labeled by $\lambda_{k}$, which, in general, have been identified in previous works~\cite{Oda13,Gom16} through higher-order recurrence relations. Therefore, the separation into independent sequences of eigenfunctions has allowed identifying simple recursion formulas for the exceptional Hermite polynomials defined by the partition $\lambda_{k}$ that, to the best of the authors' knowledge, have been unnoticed.

It is worth noting that the existence of third-order ladder operators has been one of the key properties exploited throughout the text, and thus a further generalization of ladder operators of a higher order can be considered to address more general systems. If true, the latter could lead to simple recurrence relations for a broader family of exceptional Hermite polynomials defined by more general partitions. Moreover, the potentials and solutions related to the third-order shape-invariant Hamiltonian admit general non-rational solutions that have not been exploited in detail. Those are problems that deserve special attention by themselves, and they will be discussed elsewhere.

%----------------------------------------------------

%---------Acknowledgments---------------------------
\section*{Acknowledgments}
I. Marquette was supported by Australian Research Council Future Fellowship FT180100099. V. Hussin acknowledges the support of research grants from NSERC of Canada. K. Zelaya acknowledges the support from the Mathematical Physics Laboratory of the Centre de Recherches Mat\'ematiques, through a postdoctoral fellowship. K. Zelaya also acknowledges the support of Consejo Nacional de Ciencia y Tecnolog\'ia (Mexico), grant number A1-S-24569.
%-----------------------------------------------

%---------Appendix A ---------------------------
\appendix
\setcounter{section}{0}  % reset counter 
\section{B\"acklund transformations}
\label{sec:Backlund}
\renewcommand{\thesection}{A-\arabic{section}}
% redefine the command that creates the equation no.
%\setcounter{section}{0}  % reset counter 
\renewcommand{\theequation}{A-\arabic{equation}}
% redefine the command that creates the equation no.
\setcounter{equation}{0}  % reset counter
The B\"acklund transformation is one of the most well-known techniques used to generate new solutions by departing from an already known solution, usually called \textit{seed solution}. B\"acklund transformations are given in the form of nonlinear recurrence relations. In this form, we can iterate such a transformation so that we generate different solutions at every step. Let $w_{0}\equiv w_{0}(x;\alpha_{0},\beta_{0})$ be a seed solution to~\eqref{eqw} with parameters $\alpha=\alpha_{0}$ and $\beta=\beta_{0}$, we can then generate eight new solutions through the following B\"acklund transformations~\cite{Cla08}
\begin{align}
& w_{1}^{\pm}:=\frac{\mathcal{F}_{0}^{-}\mp\sqrt{-2\beta_{0}}}{2w_{0}} \, ,  && w_{2}^{\pm}:=-\frac{\mathcal{F}_{0}^{+}\mp\sqrt{-2\beta_{0}}}{2w_{0}} \, , 
\label{w1pm}\\
& w_{3}^{\pm}:=w_{0}+\frac{2(1-\alpha_{0}\mp\frac{1}{2}\sqrt{-2\beta_{0}})w_{0}}{\mathcal{F}_{0}^{+}\pm\sqrt{-2\beta_{0}}} \, , && w_{4}^{\pm}:=w_{0}+\frac{2(1+\alpha_{0}\pm\frac{1}{2}\sqrt{-2\beta_{0}})w_{0}}{\mathcal{F}_{0}^{-}\mp\sqrt{-2\beta_{0}}} \, ,
\label{w3pm} \\
& \mathcal{F}_{0}^{\pm}:=w_{0}'\pm(2xw_{0}+w_{0}^{2}) && \, ,
\label{w00}
\end{align}
where $w_{i}^{\pm}\equiv w_{i}^{\pm}(x;\alpha_{i}^{\pm},\beta_{i}^{\pm})$ are solutions to~\eqref{eqw} with $\alpha=\alpha_{i}^{\pm}$ and $\beta=\beta_{i}^{\pm}$, for $i=1,2,3,4$. The parameters are determined in terms of $\alpha_{0}$ and $\beta_{0}$ through
\begin{align}
& \alpha_{1}^{\pm}=\frac{1}{4}\left(2-2\alpha_{0}\pm 3\sqrt{-2\beta_{0}}\right) \, , && \beta_{1}^{\pm}=-\frac{1}{2}\left(1+\alpha_{0}\pm\frac{1}{2}\sqrt{-2\beta_{0}}\right)^{2} \, , \\
& \alpha_{2}^{\pm}=-\frac{1}{4}\left(2+2\alpha_{0}\pm 3\sqrt{-2\beta_{0}}\right) \, , && \beta_{2}^{\pm}=-\frac{1}{2}\left(1-\alpha_{0}\pm\frac{1}{2}\sqrt{-2\beta_{0}}\right)^{2} \, , \\
& \alpha_{3}^{\pm}=\frac{3}{2}-\frac{\alpha_{0}}{2}\mp\frac{3}{4}\sqrt{-2\beta_{0}} \, , && \beta_{3}^{\pm}=-\frac{1}{2}\left(1-\alpha_{0}\pm\frac{1}{2}\sqrt{-2\beta_{0}}\right)^{2} \, , \\
& \alpha_{4}^{\pm}=-\frac{3}{2}-\frac{\alpha_{0}}{2}\pm\frac{3}{4}\sqrt{-2\beta_{0}} \, , && \beta_{4}^{\pm}=-\frac{1}{2}\left(-1-\alpha_{0}\pm\frac{1}{2}\sqrt{-2\beta_{0}}\right)^{2} \, .
\end{align}
In particular, if we use any of the rational solutions~\eqref{w1}-\eqref{w3} corresponding to the ``$-2x/3$'' hierarchy we obtain another rational solutions within the same hierarchy. This enables us to compute some new identities for the generalized Okamoto polynomials. 

For instance, using $w_{0}\equiv w_{m,n}^{[1]}$ into~\eqref{w1pm}-\eqref{w00}, and comparing $w_{1}^{+}$ with $w_{4}^{+}-w_{0}$, we obtain the identities
\small
\begin{align}
&-2xQ_{m+1,n}Q_{m,n+1}+3\left(Q_{m+1,n}Q'_{m,n+1}-Q'_{m+1,n}Q_{m,n+1}\right)=-\sqrt{2} Q_{m,n}Q_{m+1,n+1} \, , \\
&Q_{m+1,n+1}Q'_{m,n}-Q'_{m+1,n+1}Q_{m,n} =-\sqrt{2} (3m+3n+1) Q_{m+1,n}Q_{m,n+1} \, ,
\end{align}
\normalsize
whereas comparing $w_{2}^{+}$ with $w_{3}^{+}-w_{0}$ leads to
\small
\begin{align}
&-2xQ_{m,n+1}Q_{m,n}+3\left(Q_{m,n+1}Q'_{m,n}-Q'_{m,n+1}Q_{m,n}\right)=-\sqrt{2} Q_{m+1,n}Q_{m-1,n+1} \, , \\
&Q_{m+1,n}Q'_{m-1,n+1}-Q'_{m+1,n}Q_{m-1,n+1} =-\sqrt{2} (3m-1) Q_{m,n+1}Q_{m,n} \, .
\end{align}
\normalsize
On the other hand, after setting $w_{0}\equiv w_{m,n}^{[2]}$ and comparing $w_{1}^{+}$ with $w_{4}^{+}-w_{0}$ we obtain another set of identities of the form
\small
\begin{align}
&-2xQ_{m,n}Q_{m+1,n}+3\left(Q_{m,n}Q'_{m+1,n}-Q'_{m,n}Q_{m+1,n}\right)=-\sqrt{2} Q_{m+1,n-1}Q_{m,n+1} \, , \\
&Q_{m+1,n-1}Q'_{m,n+1}-Q'_{m+1,n-1}Q_{m,n+1} = \sqrt{2} (3n-1) Q_{m,n}Q_{m+1,n} \, .
\end{align}
\normalsize
These identities allow us to cast the rational solutions of the ``$-2x/3$'' hierarchy into the form presented in~\eqref{w1r}-\eqref{w3r}.

%--------- Appendix B ---------------------------
\appendix
\setcounter{section}{1}  % reset counter 
\section{Action of the ladder operators on the zero-modes and higher-modes}
\label{sec:APPB}
\renewcommand{\thesection}{B-\arabic{section}}
% redefine the command that creates the equation no.
%\setcounter{section}{0}  % reset counter 
\renewcommand{\theequation}{B-\arabic{equation}}
% redefine the command that creates the equation no.
\setcounter{equation}{0}  % reset counter
In this section we show that the action of $A^{\dagger}$ on the zero-modes and higher-modes indeed leads to a ladder operation. To this end, let us consider the zero-mode $\phi_{0;1}$ in terms of the Wronskian representation~\eqref{wronskian1}, together with the set indexes
\begin{equation}
\begin{aligned}
& I_{1}:=\{1,4,\cdots,3k-2;2,5,\cdots3k-1\} \, , \quad I_{2}:=\{1,4,\cdots,3k-5;2,5,\cdots3k-4\} \, .
\end{aligned}
\end{equation}
In addition, it is useful to introduce a particular identity, obtained from higher-order SUSY QM~\cite{Cru55,Mie00,Fer99}. Be $\mathfrak{H}=-\frac{d^{2}}{dx^{2}}+V(x)$ a Hamiltonian operator such that the eigenvalue equation $\mathfrak{H}\mathcal{F}_{\nu}=E_{\nu}\mathcal{F}_{\nu}$ holds for $\nu\in\mathbb{R}$, along with $E_{\nu}$ and $\mathcal{F}_{\nu}$ the eigenvalues and eigenfunctions, respectively. Then, the Wronskian identity~\cite{Oda20}
\begin{multline}
\operatorname{Wr}\left[\operatorname{Wr}(\mathcal{F}_{i_{1}},\mathcal{F}_{i_{2}},\cdots,\mathcal{F}_{i_{n}}),\operatorname{Wr}(\mathcal{F}_{i_{1}},\mathcal{F}_{i_{2}},\cdots,\mathcal{F}_{i_{n}},\mathcal{F}_{i_{n+1}},\mathcal{F}_{i_{n+2}})\right] \\
\propto \operatorname{Wr}(\mathcal{F}_{i_{1}},\cdots,\mathcal{F}_{i_{n}},\mathcal{F}_{i_{n+1}})\operatorname{Wr}(\mathcal{F}_{i_{1}},\cdots,\mathcal{F}_{i_{n}},\mathcal{F}_{i_{n+2}}) \, ,
\label{WWW}
\end{multline}
is fulfilled, where the eigenfunctions $\mathcal{F}_{i_{m}}$ are not necessarily square-integrable solutions.

The Wronskian representation of the Okamoto polynomial $Q_{k}$ can be rewritten by exploiting the properties of the functions $\psi_{n}$ given in~\eqref{psir}, such as the identity $\frac{d}{dx}\psi_{n}=\psi_{n-1}$. Moreover, we have $\psi_{0}=1$ and the Wronskian representation of $Q_{k}$ in~\eqref{wronskian1}, which leads to
\begin{multline}
Q_{k}\equiv C_{k}\mathcal{W}_{I_{2}}= C_{k}\mathcal{W}_{\{1,\cdots,3k-5;2,\cdots,3k-4\}}=C_{k}\mathcal{W}_{\{2,\cdots,3k-4;0,3,\cdots,3k-3\}}\\
=C_{k}\mathcal{W}_{\{0,3,\cdots,3k-3;1,3,\cdots,3k-2\}}=C_{k}\mathcal{W}_{\{1,\cdots,3k-2;2,\cdots,3k-1,0\}}=C_{k}\mathcal{W}_{I_{1}\cup\{0\}} \, ,
\end{multline}
where in the latter we have used the notation introduced in~\eqref{wronskiantilde0}. Thus, the zero-mode $\phi_{0;1}^{(k)}$ takes the following form
\begin{equation}
\phi_{0;1}^{(k)}=\mathcal{N}_{0;1}e^{-\frac{x^{2}}{6}}\frac{Q_{k}}{Q_{k+1}}=\mathcal{N}_{0;1}e^{-\frac{x^{2}}{6}}\frac{\mathcal{W}_{I_{1}\cup\{0\}}}{\mathcal{W}_{I_{1}}} \, .
\end{equation}
We now proceed with the action of the creation operator $A=Q^{\dagger}M_{2}M_{1}$. First, the action of $M_{1}$ is determined through
\begin{equation}
M_{1}\phi_{0;1}^{(k)}=\mathcal{N}_{1;1}\left[-\frac{2x}{3}+\frac{d}{dx}\ln\frac{\mathcal{W}_{I_{1}\cup\{0\}}}{\mathcal{W}_{I_{2}\cup\{3k-2\}}}\right]e^{-\frac{x^{2}}{6}}\frac{\mathcal{W}_{I_{1}\cup\{0\}}}{\mathcal{W}_{I_{1}}} \, ,
\end{equation}
which simplifies using the Wronskian $\tilde{\mathcal{W}}_{\{i_{1},\cdots,i_{n}\}}$, introduced in~\eqref{wronskiantilde1}-\eqref{wronskiantilde2}, resulting in
\begin{equation}
M_{1}\phi_{0;1}^{(k)}=\mathcal{N}_{1;1}\frac{d}{dx}\ln\frac{\tilde{\mathcal{W}}_{I_{1}\cup\{0\}}}{\tilde{\mathcal{W}}_{I_{2}\cup\{3k-2\}}}e^{-\frac{x^{2}}{6}}\frac{\mathcal{W}_{I_{1}\cup\{0\}}}{\mathcal{W}_{I_{1}}} \, .
\end{equation}
Then, using the identity~\eqref{WWW}, and after some straightforward calculations we finally get
\begin{equation}
M_{1}\phi_{0;1}^{(k)}=\mathcal{N}_{0;1}e^{-\frac{x^{2}}{6}}\frac{\mathcal{W}_{I_{2}\cup\{3k-2,0\}}}{\mathcal{W}_{I_{2}\cup\{3k-2\}}} \, .
\end{equation}

Following the same steps as above, the subsequent action of $M_{2}$ and $Q^{\dagger}$ is
\begin{equation}
M_{2}M_{1}\phi_{0;1}^{(k)}=\mathcal{N}_{0;1}e^{-\frac{x^{2}}{6}}\frac{\mathcal{W}_{I_{2}\cup\{0\}}}{\mathcal{W}_{I_{2}}}=\mathcal{N}_{0;1}e^{-\frac{x^{2}}{6}}\frac{\mathcal{W}_{I_{1}\cup\{3,0\}}}{\mathcal{W}_{I_{2}}} \, .
\end{equation}
and 
\begin{equation}
\phi_{1;1}^{(k)}\propto Q^{\dagger}M_{2}M_{1}\phi_{0;1}^{(k)} \propto \mathcal{N}_{0;1}e^{-\frac{x^{2}}{6}}\frac{\mathcal{W}_{I_{1}\cup\{3\}}}{\mathcal{W}_{I_{1}}} \, .
\end{equation}
respectively. The previous procedure can be easily iterated finitely many times as necessary, and after $n$ iterations we get
\begin{equation}
\phi_{n;1}^{(k)}\propto (A^{\dagger})^{n}\phi_{0;1}^{(k)}\propto e^{-\frac{x^{2}}{6}}\frac{\mathcal{W}_{I_{1}\cup\{3n\}}}{\mathcal{W}_{I_{1}}} \, .
\end{equation}

We proceed in a similar way to determine the higher-modes related to the other two sequences by starting from the zero-modes $\phi_{0;2}^{(k)}$ and $\phi_{0;3}^{(k)}$. We thus obtain
\begin{equation}
\phi_{n;2}^{(k)}\propto (A^{\dagger})^{n}\phi_{0;2}^{(k)}\propto e^{-\frac{x^{2}}{6}}\frac{\mathcal{W}_{I_{1}\cup\{3n+3k+1\}}}{\mathcal{W}_{I_{1}}} \, , \quad 
\phi_{n;3}^{(k)}\propto (A^{\dagger})^{n}\phi_{0;3}^{(k)}\propto e^{-\frac{x^{2}}{6}}\frac{\mathcal{W}_{I_{1}\cup\{3n+3k+2\}}}{\mathcal{W}_{I_{1}}} \, ,
\end{equation}
which are the higher-modes, up to a normalization constant.

\appendix
\setcounter{section}{2}  % reset counter 
\section{Higher-order SUSY transformation}
\label{sec:APPC}
\renewcommand{\thesection}{C-\arabic{section}}
% redefine the command that creates the equation no.
%\setcounter{section}{0}  % reset counter 
\renewcommand{\theequation}{C-\arabic{equation}}
% redefine the command that creates the equation no.
\setcounter{equation}{0}  % reset counter
Let us consider the dimensionless rescaled Hamiltonian for the harmonic oscillator $H_{1}$, together with the respective eigenvalue equation
\begin{equation}
\mathfrak{H}_{1}\equiv-\frac{d^{2}}{dx^{2}}+V_{1}(x) \, , \quad V_{1}(x)=\frac{x^{2}}{9}-\frac{1}{3} \, , \quad \mathfrak{H}_{1}\phi_{n}^{(1)}=E_{n}\phi_{n}^{(1)} \, ,
\label{oscH}
\end{equation}
where the eigenfunctions $\phi_{n}^{(1)}$ and eigenvalues $E^{(1)}_{n}$ are respectively given by
\begin{equation}
\phi_{n}^{(1)}(x)=\frac{e^{-\frac{x^{2}}{6}}}{\sqrt{2^{n}n!\sqrt{3\pi}}}\mathtt{H}_{n}\left(\frac{x}{\sqrt{3}}\right) \, , \quad E^{(1)}_{n}=\frac{2n}{3} \, .
\end{equation}
Now, from the iterative action of the intertwining operators $B_{j}$ and $B_{j}^{\dagger}$, constructed such that
\begin{equation}
\mathfrak{H}_{j}:=B_{j}^{\dagger}B_{j}+\epsilon_{j} \, , \quad \mathfrak{H}_{j+1}:=B_{j}B_{j}^{\dagger}+\epsilon_{j}=B^{\dagger}_{j+1}B_{j}+\epsilon_{j+1} \, ,
\end{equation}
with $\epsilon_{j}$ a real constant, known as the \textit{factorization energy}, and the Hamiltonians of the form
\begin{equation}
\mathfrak{H}_{j}:=-\frac{d^{2}}{dx^{2}}+V_{j}(x) \, ,
\end{equation}
where the intertwining are operators defined as
\begin{equation}
B_{j}:=\frac{d}{dx}+\beta_{j}(x) \, , \quad B_{j}^{\dagger}:=-\frac{d}{dx}+\beta_{j}(x) \, ,
\label{Bj}
\end{equation}
and the superpotentials $\beta_{j}$ are determined from the Riccati equation
\begin{equation}
-[\beta_{j}]'+[\beta_{j}]^{2}=V_{j}-\epsilon_{j} \, , \quad [\beta_{j}]'+[\beta_{j}]^{2}=V_{j+1}-\epsilon_{j} \, , \quad V_{0}=\frac{x^{2}}{9} \, .
\end{equation}
The new potential partners are 
\begin{equation}
V_{j+1}(x)=\frac{x^{2}}{9}+2\frac{d}{dx}\sum_{k=1}^{j}\beta_{k} \, .
\label{potVj}
\end{equation}

Following the conventional linearization procedure of the Riccati equation, it is straightforward to obtain the set of linear equations
\begin{equation}
\beta_{r}^{(p)}:=-\frac{d}{dx}\ln u_{r}^{(p)} \, , \quad -\frac{d^{2}}{dx^{2}}u_{r}^{(p)}+V_{p}(x)u_{r}^{(p)}=\epsilon_{r}u_{r}^{(p)} \, ,
\end{equation}
with $u_{r}^{(p)}$ the seed functions associated with the factorization energy $\epsilon_{r}$ of the $p$-th partne potential. The superpotentials in~\eqref{Bj} correspond to the particular case $\beta_{r}^{(r)}\equiv\beta_{r}$. From the latter, we can write the $j$-th partner potential $V_{j+1}(x)$ in term of the eigenfunctions $\Phi_{n}\equiv\phi_{n}^{(1)}$ of the initial Hamiltonian $\mathfrak{h}_{1}$ as
\begin{equation}
V_{j+1}(x)=\frac{x^{2}}{9}-2\frac{d^{2}}{dx^{2}}\ln \mathcal{W}(\Phi_{\epsilon_{1}},\Phi_{\epsilon_{2}}\cdots\Phi_{\epsilon_{j}}) \, , \quad \Phi_{\epsilon_{j}}(x)\equiv \phi_{\epsilon_{j}}^{(1)}(x) \, .
\label{eq:potVj}
\end{equation}

%---------------------------------------> Bibliography

\end{document}